# Effects of Particle sizes, Non-Isometry and Interactions in Compressible Polymer Mixtures

P. D. Gujrati

The Department of Physics, and The Department of Polymer Science
The University of Akron, Akron, OH 44325, USA

**Abstract**

We consider in this review the statistical mechanical description of a very general microscopic lattice model of a compressible and interacting multi-component mixture of linear polymers of fixed lengths. The model contains several microscopic, i.e. bare parameters determining the thermodynamic state of the system. General arguments are given to show that these parameters must be independent not only of the lattice properties but also of the thermodynamic state, and that the voids representing free volume must be carefully treated, if thermodynamics has to be properly obeyed. These facts have not always been appreciated in the literature. We focus on mixing functions, some of which have not been properly calculated in the literature. In general, mixing is non-isometric (volume of mixing $\Delta_{\mathrm{M}} V \neq 0$) and the entropy of mixing is non-ideal. We have recently developed a lattice theory for the general model, which goes beyond the random mixing approximation (RMA) limit and is thermodynamically consistent in the entire parameter space. The theory contains terms that do not have a continuum analog except in the RMA limit or for point-like particles. Both the free volume and the total volume determine the thermodynamics of the system. The RMA limit of our theory gives rise to a new theory, which can be taken as the extension of the conventional incompressible Flory-Huggins theory and is similar in simplicity. Using our complete theory, we calculate the effects of size disparity and interactions on the thermodynamics of the model. Cohesive energies are not constant in general. Non-isometry can make the energy of mixing negative, even when all exchange interactions are repulsive. Consequently, Scatchard-Hildebrand theory cannot be substantiated in general. Various unusual features are noted and discussed.

## I.   Introduction

This review deals with the application of classical statistical mechanics[1-3] to study multi-component polymer mixtures using first principles, rather than phenomenology. This requires introducing *microscopic* or *bare* parameters at the outset in an appropriate model of the system, in terms of which thermodynamic quantities are to be expressed. Thus, obtaining the values of the bare parameters experimentally or by first principles is a challenging endeavor. The general model should describe not only the mixture, but also the pure components in certain limits. The role of statistical mechanics, to a good part, in the theory of mixtures is to provide us with a microscopic description of mixtures, and to enable us to calculate changes that occur when fluids are mixed [4-8] in terms of these bare parameters. *The description will certainly be thermodynamically consistent if the calculations are done exactly.* An exact calculation will enable us to extract the precise values of the microscopic parameters from the measurement of thermodynamic quantities like the interaction energy, the compressibility, the volume of mixing, etc. Of course, we are implicitly assuming here that the complete model correctly represents the real system. Unfortunately, it is a reality that no exact statistical mechanical calculations can be carried out at present for the complete model of the mixture. To make further progress, *approximations* have to be made. The results of these calculations give rise to various (approximate) theories of the same model and each of which allows us to *estimate* the microscopic parameters. The estimates, known as *effective* or (thermodynamically) *dressed* parameters, depend obviously on the theory employed and are usually imprecise. In addition, in several theories, the thermodynamic consistency of the approximations *cannot* be established, especially when the approximations are *phenomenological* in nature. Indeed, it has been known for quite sometime[4] that the Percus-Yevick approximation in continuum yields the pressure and the compressibility equations that are mutually inconsistent; in addition, it also gives negative probabilities and, hence, physically unacceptable solution for high packing densities.[9] For this and various other reasons (see below), we are only interested in lattice theories in this review; moreover, we will be interested primarily in theories based on first principles, rather than phenomenology with special emphasis on their comparison.

If one can establish that a certain approximation is mathematically equivalent to solving the model *exactly* on some special lattice, no matter how unrealistic the lattice, it automatically *ensures* that *the approximation is thermodynamically consistent*. The solution of the model on the special lattice gives rise to an *approximate* but *consistent theory* of the model on the original lattice. The usefulness of the special lattice and the corresponding approximate theory is then determined by how useful the predictions of the theory are for real systems it proposes to describe. For example, the conventional mean-field theories like the ideal or the regular solution theories are based on random mixing approximation (RMA), and can be shown to be *exact* on an infinite coordinated lattice, or on an equivalent-neighbor lattice on which each particle is a neighbor of all other particles.[2,3] Even though both lattices are extremely unrealistic representations of a real system, their importance cannot be overstated. The above-mentioned exactness of this calculation automatically ensures that these theories are thermodynamically consistent. Moreover, they are microscopic theories based on first principles. Because of this, these theories have played, and will continue to play, an important role in developing thermodynamics. However, when one tries to fit experimental results to these theories, they turn into phenomenological theories in that the parameters in the theories no longer remain microscopic in nature. As a consequence, they become a function of the thermodynamic state of the system, which leads to certain thermodynamic inconsistencies, as the review will show.

(i) **Ideal and Regular Solutions, and RMA**

One of the conditions for an ideal or a regular solution is that mixing be *isometric*, i.e. $\Delta_M V = 0$ not only in that state, but also in all states nearby. We use isometric in this sense here. The regular solution theory is mathematically equivalent to the random-mixing approximation (RMA) for an incompressible mixture.[2] The *incompressible RMA limit* requires $q\to\infty$, $T\to\infty$, such that $q/T$ remains fixed and finite.[10-13] Here, $q$ is the coordination number of the lattice, and the temperature $T$ is measured in the units of the Boltzmann constant $k_B$, i.e. we set $k_B=1$. The RMA limit for a compressible system is easily identified by additionally requiring $P\to\infty$ such that $P/T$ is fixed and finite.[14] We have already shown[12] that a theory in the *RMA limit* can be constructed for a *compressible* lattice model. The compressibility is obtained by introducing voids.[15,16] The compressible RMA theory is *not* a regular solution theory since $\Delta_M V \neq 0$ in general, because of asymmetry in interactions and/or particle size difference.[12]

Generally, the regular solution theory is developed on a lattice by considering an *incompressible* mixture of particles so that $\Delta_M V=0$. The smallest inter-particle distance between particles of any species is taken to be the same and equal to the lattice spacing $a_0$; see Fig. 1(a) where the centers, shown by filled dots, of particles, shown by thick squares, of different sizes are separated by one lattice spacing. We call this to be the principle of isometry. The actual particle sizes play no important role in the calculation as long as the above condition is satisfied. Thus, it is many times convenient to take each particle to have the *same* volume equal to the cell volume $v_0 = a_0^d$ ($d$ being the lattice dimension), shown by broken squares in Fig. 1(a), regardless of the species. We say that all particles have the same size in this sense in the following. The mixing remains isometric for such particles even in a compressible lattice model. Voids, each of which also has the same size $v_0$, are introduced for the compressibility. In both cases, the entropy calculation in the *athermal* state is trivial.[6,8,12] The problem arises when particles are *different* in sizes such as in a polymer solution. Fowler and Rushbrooke[17] have already argued that an athermal solution of dimers and monomers will not be an ideal solution. While the shortest distance between two monomers or two parallel dimers is $a_0$, it is $\sqrt{5}\,a_0/2$ between a dimer and a monomer at right angle; see Fig. 1(b). Consequently, the athermal mixing becomes non-isometric. More recently, Gujrati[12] has shown that, in general, $\Delta_M V \neq 0$ due to size disparity in an athermal mixture. The presence of interaction gives rise to $\Delta_M V \neq 0$ even if there is no size disparity.

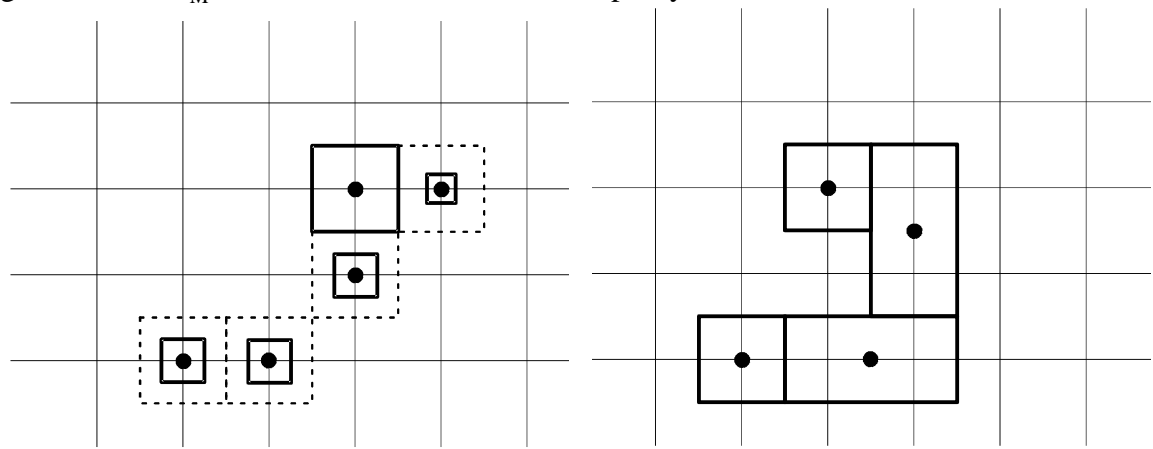

Fig. 1. (a) Isometric and (b) non-isometric mixtures. The particle centers of mass (COM's) are shown by filled dots.

The situation is not very clear in continuum theories. The problem of particles with a non-zero size in continuum is still an outstanding unsolved problem, though major progress has been made.[7] We make a few comments below and refer the reader to Refs. 4-7 for details. There is a striking *similarity* in the form of the *ideal* entropy of mixing $\Delta_M S$ in a continuum theory and its lattice version. Despite this, there are major differences. The continuum approach treats the system as *compressible*, with particle sizes never appearing in the deliberation. The lattice theory obeying the principle of isometry is for an *incompressible* system for particles with real sizes less than or equal to $v_0$. As we will argue below, the simple-minded attempts to calculate $\Delta_M S$ in continuum are only for *point-like* particles, so that the ideal gas equation is obeyed. As discussed in detail in Ref. 12 and in Sect. II here, the extension to particles with non-zero sizes in the spirit of van der Waals approach gives *P* that diverges as a power law in the free volume and is different from the logarithmic divergence in *P* in the lattice formulation. In addition, we will show below that continuum theories á la van der Waals are not very encouraging.[4-7] Therefore, we are again forced to limit our deliberation to the *lattice* formulation here. We show in this review that many of the mixing functions have been calculated incorrectly in some theories. The correct procedure for calculating them is described here.

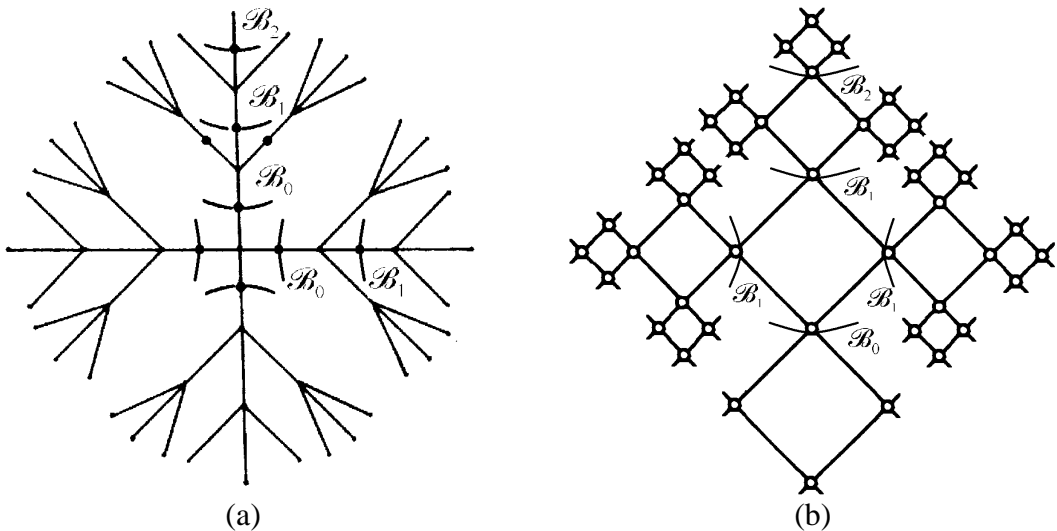

(a)          (b)
Fig.2. (a) Bethe lattice. (b) Husimi cactus.

**(ii)    Recursive Lattice Approach**

In general, real systems do not correspond to *q*, *P* or *T*, all diverging to infinity. Therefore, we need to consider a theory that goes beyond the RMA. To accomplish this goal, we have adopted the above-mentioned approach of approximating the original lattice by a *special* lattice on which the statistical mechanical model is solved *exactly*. In our recent investigations, we have taken for the approximate lattice a recursive lattice. Such lattices are amenable to exact calculations. A recursive lattice is a structure obtained recursively in terms of its smaller parts.[10,11] For example, the Bethe lattice in Fig. 2(a) has the property that each of the four main branches $B_0$ is made up of three smaller branches $B_1$, each of which is made up of still three smaller branches $B_2$, and so on. The entire lattice is infinitely large, and we show only a small part of it. A small part of another recursive lattice, known as the Husimi cactus lattice, is shown in Fig. 2(b) along with its various branches. Due to the recursive nature, the model can be solved *exactly* without any further approximation. The exactness of the solution implies that the resulting theory is thermodynamically consistent, and is taken as the *approximate* theory of the model on the original lattice. This theory goes beyond the RMA limit, and gives corrections to any thermodynamic quantity over its value in the regular or ideal solution theory.

The two recursive lattices in Fig. 2 form possible approximations for a square lattice; all three lattices have the same coordination number $q=4$. The Bethe lattice in Fig. 2(a) should be more appropriate for describing the physics on a square lattice in contrast with an infinite-$q$ lattice used in the RMA. However, the former has no closed loop, so it misses some important correlations induced by these loops on a square lattice. This is remedied in the Husimi cactus in Fig. 2(b), which contains the smallest possible loops that are present on a square lattice. Thus, the solution on the Husimi cactus will be even better than the one on the Bethe lattice in Fig. 2(a) for a square lattice. A proper choice of a recursive lattice, which incorporates correlations that are lost in the RMA, and which allows for exact calculation will give rise to a theory not only superior to a RMA-based theory but will also be thermodynamically consistent. We have already demonstrated this elsewhere.[18] We have had tremendous success not only in investigating the bulk behavior, but also in confined geometries over the past decade. We refer the reader to two recent reviews[10,11] of this approach.

**(iii)   Scope of the Review**

The most important goal of the review is to highlight the independence of bare parameters on the lattice properties and on the thermodynamic state of the system, as it has not been properly appreciated in the literature. We also illustrate the consequences of our new theory of compressible mixtures, as the two recent reviews[10,11] do not specifically deal with it. The properties of mixing can be calculated for a variety of mixing processes, which can be carried out at constant pressure ($P$), at constant volume ($V$), at constant temperature ($T$), etc. However, for liquids, the most commonly studied process of mixing is at constant $T$ and $P$. It is this process that we will consider mostly in this review. The use of various thermodynamic quantities of mixing in the regular solution theory has been a standard practice since the nineteen thirties. The chief reason for considering mixing functions is that it allows us to exploit the arbitrariness in the definition of the functions that does not affect the thermodynamics of phenomena like phase separation. For example, the entropy $S$ in a continuum model can become infinitely large in many cases; therefore, it is usually defined up to a constant and the constant cannot affect the thermodynamics of the system. Similarly, the chemical potential is also defined up to a function of temperature only. Such arbitrariness plays no useful purpose. In this review, we are mainly interested in the mixing functions of extensive quantities $V, S, E, G$, etc. that represent the total volume, the total entropy, the total internal energy, the total Gibbs free energy, etc. respectively. Mixing functions $\Delta_M V, \Delta_M S, \Delta_M E, \Delta_M G$, etc. allow us to exploit this arbitrariness in a simple manner and, therefore, play a very constructive role in statistical mechanics and thermodynamics of mixtures. We are also interested in mixing quantities per monomer denoted by lower case letters: $\Delta_M v_m, \Delta_M s_m, \Delta_M e_m, \Delta_M g_m$, etc. The additional subscript "m" is used to distinguish them from quantities that are defined per lattice site. For the mixing function, one subtracts from the extensive quantity $Q$ its value in some suitable *reference* state. Conventionally, one takes the unmixed, pure state of the system as the reference state. In that case, the thermodynamic quantity $\Delta_M Q$ of mixing is the difference between the quantity $Q$ and the sum of this quantity for the same amount of pure components at the same *temperature T* and the *pressure P*. It is obvious that $\Delta_M Q$ vanishes in the reference state. However, as we will see below, there are conflicting results for mixing quantities in lattice theories. We hope that the review will settle the confusion.

It should be stressed at this point that the usefulness of mixing functions lies in the observation that the thermodynamics of the mixture is *not* affected by the choice of the reference state. For this to be true, the reference state must be *independent* of the mixture state. It is convenient in some cases to use a reference state that may not even exist in Nature. In addition, one can add a constant or a function linear in densities to a mixing quantity without affection the phase diagram. Therefore, the choice of a pure system at the same temperature and pressure as the mixture for the reference system is merely a convention. It is a common practice to impose certain *ad hoc* mixing rules to express mixture parameters in terms of pure system parameters. We will *not* impose any *ad hoc* mixing rules on the system in our approach. Instead, we will introduce mixture parameters in the form of microscopic bare parameters in the statistical mechanical description, which can later be determined or estimated directly from experiments.

We will demonstrate that it is, in general, *impossible* to cast results from a lattice model in a form that will be meaningful for a continuum description. In fact, the lattice formulation depends on the lattice coordination number, which has no counterpart in continuum. The two formulations appear to have the same *form* only in the RMA limit or for point-like particles. Since such limits are *unrealistic*, we conclude that establishing any similarity of form between lattice and continuum formulation is *not* possible in general. Another important distinction between the lattice and continuum formulation is presented in Ref. 14, where it has been argued that the microscopic exchange energies can be sensibly defined only in a lattice model, but not in a continuum model. This result is a consequence of the fact that the lattice used in a lattice formulation remains fixed for all configurations of the system, while any appropriately defined lattice structure in continuum keeps changing with the configuration in continuum.

The layout of the review is as follows. In the next section, we consider athermal and incompressible simple and polymer mixtures systems on a lattice to calculate $\Delta_M S$ in the RMA limit. We contrast lattice and continuum approaches needed to calculate $\Delta_M S$, and discuss the limitations of the continuum approach. We also consider two possible extensions of the traditional Flory-Huggins (F-H) theory to a compressible system that are currently used in the literature. We demonstrate that neither of them can be justified as proper. We give in Sec. III a general description of the statistical mechanics of a compressible multi-component system on a lattice. We argue that voids must be treated as a special kind of species than material species, which constitute our physical systems. The following section (IV) deals with some important cautionary remarks regarding lattice properties and system parameters that seem to have been misunderstood in the literature. In Sect. V, we describe without giving much details the recursive lattice theory that has been recently developed by us. This theory is used to explicitly calculate various mixing functions and related quantities like the effective chi, the cohesive energy density, etc. in the remainder of the review. We show in this section (V) that our theory in the RMA limit gives rise to a *new* theory of compressible polymer mixture that is at the same level of approximation as the original incompressible F-H theory. The following four sections deal with the calculation of the volume of mixing, the entropy of mixing, the energy of mixing, and the Gibbs free energy of mixing, respectively. We also discuss these quantities in the RMA limit when appropriate. We discuss the cohesive energy density and the solubility parameter and their relationship with an effective chi that is directly related to the energy of mixing. We present some numerical results in Sect. X. The final section (XI) contains conclusions and a brief summary of the results. The three appendices contain some technical proofs of results presented in the main text.

## II. Lattice Formulation: Athermal Mixtures

**(a) Simple Mixtures.** The *ideal* entropy of mixing for a simple binary incompressible solution of $N_1$ and $N_2$ particles of species 1 and 2 on a *lattice* and satisfying the principle of isometry is[19]

$$\Delta_M S = - N_1 \ln y_1 - N_2 \ln y_2, \qquad (1)$$

where $y_1 \equiv N_1/N$, $y_2 \equiv N_2/N$ are *number fractions*, and $N \equiv N_1 + N_2$ is the number of lattice sites, so that $y_1 + y_2 = 1$. The centers of the particles reside on lattice sites and no site is occupied by more than one particle. We can take the size of each particle to be $v_0$, even if the real size is less than or equal to $v_0$, if the principle of isometry is obeyed.

In a continuum model, the entropy of mixing can also be easily calculated provided we treat the particles as *point-like*, so that $\Delta_M V = 0$. Let $V'$, $V''$ denote the volumes of the two pure components of species 1, 2, respectively, and $V$ that of the mixture. Let $v' \equiv V'/V$, and $v'' \equiv V''/V$ be the *volume ratios*. For $\Delta_M V = 0$, $V = V' + V''$, and the ratios become fractions, which add up to one. In this case, the probability of finding a particle of species 1 (or 2) in a volume $V'$ (or $V''$) is $v'$ (or $v''$). Thus,[1]

$$\Delta_M S = - N_1 \ln v' - N_2 \ln v''. \qquad (2)$$

Using the ideal gas equation $PV = N_n T$, $N_n$ being the number of particles, to the two pure components, we can transform Eq. (2) into Eq. (1). This is a *remarkable result* showing the *equivalence* of lattice and continuum versions for the ideal entropy of mixing for athermal particles at constant $T$ and $P$. However, it says nothing about the equality of the entropies themselves. It should also be stressed that Eq. (2) is only *valid* for point-like particles, while the particles in the lattice model can be considered to have a size less than or equal to the cell size.

One purpose of the lattice is to ensure that the shortest distance between the centers of mass of two particles is a lattice bond length so that the principle of isometry is obeyed. In this interpretation, which in some sense may be more appealing, the size of the particle is no longer a useful parameter as long as it is less than the lattice cell size, and may very well be taken to be zero such as a for a point. The interpretation, though somewhat unconventional, brings out the above similarity clearly. However, there is another very important aspect of a lattice model that is usually not appreciated by workers in the field. The presence of a finite and non-zero *lattice bond* size provides a *short distance cut-off* so that the entropy per site in *any* state remains *bounded*. This is similar to providing a finite and non-zero cell size in the configuration space due to the Planck's constant $\hbar$. Thus, both continuum and lattice models contain a cut-off. In the lattice model, the cut-off may or may not be related to the particle sizes. As we will see, its presence only affects the definition of the volume and the pressure, but not the entropy. In a continuum model, the Planck's constant only affects the entropy, but not the pressure. As is well known, the quantum cut-off ensures that the entropy remains finite (and positive) in all cases, whereas it can become infinitely large and negative in classical statistical mechanics.[1,2] As we will see below, continuum models indeed give rise to such a divergence, but not lattice models. This is one reason why the (ideal) entropy of mixing is such a useful quantity in the continuum approach, as it does not diverge.

   **(b) Flory-Huggins (F-H) Theory.** Huggins[20] and Flory[21,22] independently calculated the F-H entropy of mixing of long chain molecules on a *lattice* in the *incompressible* limit ($\Delta_M V = 0$), which is identical to Eq. (1), except that the number fractions are replaced by the *monomer fractions* $y_1 \equiv N_1 M_1 / N_m$, $y_2 \equiv N_2 M_2 / N_m$, respectively. Here, $M_1$ and $M_2$ are the numbers of monomers in each polymer chain and $N_1$ and $N_2$ are the numbers of polymer chains of species $j = 1$ and 2, respectively, and $N_m \equiv N_1 M_1 + N_2 M_2$ is the total number of monomers in all polymers and also the number of sites in the lattice.[23] The same result is also valid in continuum, as shown by Longuet-Higgins,[24] provided we assume $\Delta_M V = 0$ and make an assumption regarding the two-body correlation functions that is equivalent to assuming point-like particles. Thus, even for polymer chains, the lattice and continuum calculations of $\Delta_M S$ give identical results, provided $\Delta_M V = 0$ and the particles are point-like in the continuum version. In the lattice model, we may adopt any of the two interpretations noted earlier. Again, the entropies themselves will not be identical.

   **(c) Hildebrand & van der Waals Theories.** Hildebrand[6,25] has argued that when particles are *not* point-like, $\Delta_M S$ is given by

$$\Delta_M S = - N_1 \ln(V_0'/V_0) - N_2 \ln(V_0''/V_0), \qquad (3)$$

where the $V_0$, $V_0'$ and $V_0''$ are the *free volumes* of the mixture and the two pure components, respectively, and is said to be applicable[6] even if $\Delta_M V \neq 0$. The free volume is obtained by subtracting the volume of all the particles in the (mixture or the pure) system from the volume of the system. It is important to note that the above entropy does *not* depend on the volume itself, but only on the free volume, and is very *different* from the *ideal* $\Delta_M S$. For point-like particles, Eq. (3) reduces to Eq. (2).

   Two interesting consequences of Eq. (3) should be noted, the first of which has been discussed in details in Ref. 12. (i) The entropy of expansion of $N_n$ particles from a state with free volume $V_0'$ to a state with free volume $V_0$ is given by $\Delta_{exp} S = - N_n \ln(V_0'/V_0)$. This entropy of expansion is also valid for a one-component van der Waals fluid.[25] It becomes infinitely large, if the original state is an incompressible state with zero free volume. The divergence is due to the power-law behavior of $P$ in terms of $V_0$ and is a manifestation of non-zero particle size. Such a divergence never arises in a lattice theory.[12] (ii) For the athermal case, the van der Waals equation of state reduces to $PV_0 = N_n T$, which is similar to the ideal equation of state, except that the total volume is replaced by the free volume. Since the mixing is carried out at constant $T$ and $P$, we find that the arguments of the logarithms in Eq. (3) reduce to number fractions and we retrieve Eq. (1). In this case, there *cannot* be any volume of mixing, even if the two particles have *different* size. Thus, the van der Waals equation does *not* explain the phenomenon noted above of a non-zero volume of mixing for particles of different sizes.

**(d) Extending F-H Theory to Compressible Systems**. There are at least two *different* extensions of the F-H entropy of mixing to the compressible lattice system that are used by various authors. The first one is identical in form to Eq. (2), except that the arguments for the logarithms are *volume fractions* and not monomer fractions, so that the volume of mixing is neglected.[26-34] Unfortunately, this formulation does not properly account for the entropy due to free volume; see Eq. (3). Problems related to this approach are discussed by us elsewhere,[14] and we refer the reader to this work for further details.

The other extension is to modify the F-H entropy of mixing by adding to it, and not to the total entropy, the contribution from the free volume. The free volume is modeled[12,35-37] by a new species ($j = 0$) called voids or holes.[15,16] Each void has the *same* size $v_0$ as other monomers and occupies a site of the lattice. No site is occupied more than once. The entropy of mixing is[38-41] taken to be

$$\Delta_M S = -N_0 \ln \phi_0 - N_1 \ln \phi_{m1} - N_2 \ln \phi_{m2}, \quad (4)$$

where $N_0$ is the number of voids, $N \equiv N_0 + N_1 M_1 + N_2 M_2$ the number of lattice sites, and

$$\phi_0 \equiv N_0/N, \; \phi_{m1} \equiv N_{m1}/N, \; \phi_{m2} \equiv N_{m2}/N, \quad (5)$$

the densities of voids and monomers, respectively; $N_{m1} \equiv N_1 M_1$, and $N_{m2} \equiv N_2 M_2$.

Sanchez and coworkers[35,40,41] have justified Eq. (4) by mixing three very special "pure components", each denoted by a superscript $j$. Two of the components are the incompressible polymer systems ($P \to \infty$) corresponding to $j = 1$ and 2. The third component ($j = 0$) is a "system of voids," i.e., a pure vacuum containing $N_0$ voids and obviously corresponds $P = 0$. Thus, they define

$$\Delta_M S \equiv S(T, P, N_0, N_1, N_2) - S^{(1)}(T, \infty, 0, N_1, 0) - S^{(2)}(T, \infty, 0, 0, N_2) \\ - S^{(0)}(T, 0, N_0, 0, 0). \quad (6)$$

However, $\Delta_M S$ in Eq. (4) cannot be the *correct* entropy of mixing, as it does not vanish when $N_1=0$ or $N_2=0$ at the given $T$ and $P$. The mixing process is isometric ($\Delta_M V = 0$) but *not* isobaric. Also, the change in $PV$ cannot be calculated. As a consequence of this mixing process, and various mixing rules, other somewhat more serious problems appear in their theory; see Sect. IV, and Eq. (39) along with the discussion following it. Sanchez and Panayiotou[41] give the expression for the Gibbs free energy of mixture for lattice and continuum versions, separately. However, for the continuum version, see Eq. (92) in Ref. 41, they use the entropy of mixing and not the entropy of the mixture.

## III. Statistical Mechanics of voids as a Special Species

We consider a lattice model of a multi-component mixture of different species indexed by $j \geq 0$. Species with $j \geq 1$ are called "material" species to distinguish them from voids or holes for which we reserve the index $j = 0$. The interactions are restricted between pairs of monomers of different species located on nearest-neighbor sites. Thus, the coordination number $q$ of the lattice appears explicitly in the energy of a configuration. The lattice to be used for the mixture is a *fixed* structure, independent of the thermodynamic state of the system (composition, $T$ and $P$, etc.), and *is* characterized below by two of its important characteristics, $q$ and the cell volume $v_0$. The *connectivity* of the lattice is just as important; however, it is hard to characterize it in a simple manner so we avoid its use to characterize the lattice.

It is customary to treat the number of lattice sites $N$ as an *extensive* quantity that *is* kept *fixed* in order to obtain the *thermodynamic limit*. We denote the partition function for such a lattice by $Z_N(\bullet)$; the filled dot ($\bullet$) denotes the set of independent arguments that are kept fixed. The sequence of the "finite lattice free energy" defined by $\omega_N(\bullet) \equiv (1/N)\ln Z_N(\bullet)$ is expected to converge to the thermodynamic "free energy" or "potential" $\omega(\bullet)$, as $N \to \infty$. Each polymer of species $j$ contains $M_j \geq 1$ monomers, known as its degree of polymerization (DP). For solvent species or voids, $M_j = 1$. Furthermore, $N_{mi} = N_i M_i$ denotes the total number of monomers (m is used to imply monomers) of the $i$-th species, and $N_{ij}$ the number of nearest-neighbor contacts between monomers of species $i$, and $j \geq i$. The sum of *all* monomers and solvent particles, and the voids ($j \geq 0$) must add up to $N$: $N = N_0 + \sum_{j \geq 1} N_j M_j$. Because of this, we take the void number $N_0$ to be the dependent quantity.[10,11,12,14,42] The total and the free volumes are given by $V \equiv N v_0$, and $V_0 \equiv N_0 v_0$, respectively. The partition function is given by

$$Z_N(\beta, \{\mu_{mj}\}, \{\varepsilon_{ij}\} | q, \{M_i\}) \equiv \sum_{\Gamma:\{N_{mi}\},\{N_{ij}\}} \left( \prod_{i \geq 1} (K_{mi})^{N_{mi}} \right) \left( \prod_{(ij)} (w_{ij})^{N_{ij}} \right), \qquad (7)$$

where the sum is over distinct configurations $\Gamma$ of polymers consistent with a fixed $N$. The first product is over $i \geq 1$ and the second product is over distinct pairs ($ij$) of different species including $j = 0$. The $j$-th species ($j>0$) monomer activity is $K_{mj} \equiv \exp(\beta \mu_{mj})$, where $\mu_{mj}$ is the chemical potential per $j$-th species monomer. Similarly, the Bolzmann weight $w_{ij} \equiv \exp(-\beta \varepsilon_{ij})$, where $\varepsilon_{ij}$ is the bare exchange energy between the pair ($ij$) and is given by

$$\varepsilon_{ij} \equiv e_{ij} - (e_{ii} + e_{jj})/2, \qquad i,j \geq 0, \qquad (8)$$

in terms of the pair-wise bare interaction energy $e_{jj}$. For $i = 0$, we have $\varepsilon_{0j} = -e_{jj}/2$.

As said earlier and shown elsewhere,[14] the energy combinations $\varepsilon_{ij}$ are meaningful only on a lattice, whether homogeneous or inhomogeneous, but not in a continuum. The corresponding bare chi parameters are conventionally defined by the adimensional combination $\chi_{ij} \equiv q\beta\varepsilon_{ij}$. The coordination number $q$ has been absorbed for convenience. It should also be noted that the energy parameters $\varepsilon_{ij}$, and the chemical potentials $\mu_{ij}$ must be local quantities, independent of the thermodynamic state of the system, as shown in the Appendices I, and II. Similarly, the lattice parameter $q$, and the DP $M_i$ must also be independent of the thermodynamic state. Consequently, these microscopic parameters must be, for example, composition-independent. Determination of any composition-dependence requires averaging over a part of the lattice and, in case there is inhomogeneity due to coexistence, there will be ambiguity as to what composition has to be taken. The parameters must also be independent of $T$, and $P$; otherwise we violate thermodynamic relations, as shown in the next section and in the Appendices I, and II.

The total entropy $S_N(\{N_{mi}\},\{N_{ij}\}|q,\{M_i\})$, which is the logarithm of the number of distinct configurations for given sets $\{N_{mi}\}$, and $\{N_{ij}\}$, consistent with the lattice size $N$, is a function only of $N$, $\{N_{mi}\}$, and $\{N_{ij}\}$. In addition, it also depends on $q$, and the set $\{M_i\}$. In the thermodynamic limit $N \to \infty$ such that various extensive quantities have well-defined densities per lattice site (see Refs. 10-14, and 42 for details), the entropy per site $s_N \equiv S_N/N$ reaches its limit $s$; the limit is a function of the sets of monomer densities $\{\phi_{mi}\}$ and of the contact densities $\{\phi_{ij}\}$ normalized per site, respectively, where $\phi_{mi} \equiv N_{mi}/N$, and $\phi_{ij} \equiv N_{ij}/N$ in the thermodynamic limit. The free energy $\omega_N(T,\{\mu_{mi}\},\{\varepsilon_{ij}\}|q,\{M_i\})$ has the limit $\omega$, which depends only on field variables [the temperature, the sets $\{\mu_{ij}\}$, and $\{\varepsilon_{ij}\}$], in addition to $q$, and the set $\{M_i\}$. However, it is important to note that both $\omega$, and $s$ do *not* depend on $v_0$, as the latter does not appear in Eq. (7). Because of this, $s$ can have explicit dependence only on densities defined per site, but not on densities per unit volume. In equilibrium, $\omega$, and $s$ are related to each other by the Legendre transform

$$\omega(\beta,\{\varepsilon_{ij}\},\{\mu_{mi}\}) = s(\{\phi_{mi}\},\{\phi_{ij}\}) - \beta\sum_{ij}\varepsilon_{ij}\phi_{ij} + \beta\sum_{i}\mu_{mi}\phi_{mi} , \qquad (9)$$

so that $\omega$ is no longer a function of the densities. Mathematically, this requires varying the densities on the right-hand side of Eq. (9) so as to *maximize* the free energy. Thus, the conditions to achieve equilibrium are

$$(\partial\omega/\partial\phi_{mi}) = 0, (\partial\omega/\partial\phi_{ij}) = 0 . \qquad (10)$$

The derivatives in Eq. (10) are taken at *fixed* fields $\beta$, $\{\varepsilon_{ij}\}$ and $\{\mu_{mi}\}$, and all remaining densities not involved in the differentiation. Using Eq. (9) in Eq. (10), we immediately obtain the following fundamental thermodynamic relations in terms of the entropy

$$(\partial s/\partial\phi_{mi}) = -\beta\mu_{mi}, \quad (\partial s/\partial\phi_{ij}) = -\beta\varepsilon_{ij} . \qquad (11a)$$

In terms of total entropy, the above can be rewritten as

$$(\partial S/\partial N_{mi}) = -\beta\mu_{mi}, \quad (\partial S/\partial N_{ij}) = -\beta\varepsilon_{ij} . \qquad (11b)$$

The two sums (the second and the third terms) in Eq. (9) represent the adimensional energy, and the Gibbs free energy. Hence, we conclude[12] that the free energy $\omega$ in Eq. (9) represents the adimensional pressure $z_0 \equiv \beta P v_0 = \beta\tilde{P}$; here $\tilde{P} \equiv Pv_0$ is a redefined "work"-like field variable. At coexistence, we must have the equality of $z_0$ in the coexisting phases. Since the temperature must be the same in the coexisting phases, we must require the equality of $\tilde{P}$ in the coexisting phases. Assuming $v_0$ to be a constant, this immediately leads to the equality of the pressure $P$ in the coexisting phases. Thermodynamic equilibrium [maximization of $\omega$; see Eq. (10)] requires that $z_0$ must be a maximum. In addition, since the partition function in Eq. (7) contains the vacuum state, which contributes 1 to the sum, it is evident that $z_0$ must be non-negative.[12] The constancy of $v_0$ then implies that the pressure must not only be *positive*, but must have its maximum possible value in equilibrium. On the other hand, the pressure can become *negative* in a *metastable* state.[43]

A pure component has the property that its pressure can take any possible value. The vacuum *cannot* be treated as a pure component, since its pressure is always zero and cannot be varied. Let us consider a pure component of a single material species *j*. Let the corresponding lattice be characterized by its coordination number $q^{(j)}$, and the cell volume $v_0^{(j)}$. As we will see immediately in the following section, we are forced to use the same lattice for all pure components as for the mixture. The total number of lattice sites $N^{(j)}$ must be equal to the sum of the void number $N_0^{(j)}$ and the number of monomers $N_{mj}$. Let $N_{0j}^{(j)}$ denote the number of nearest-neighbor contacts between the voids and monomers. The corresponding partition function is given by

$$Z_{N^{(j)}}^{(j)}(\beta,\mu_{mj},\varepsilon_{0j}|q^{(j)},M_j) \equiv \sum_{\Gamma:N_{mj},N_{0j}^{(j)}} [K_{mj}^{(j)}]^{N_{mj}} w_{0j}^{N_{0j}^{(j)}}. \qquad (12)$$

Here, $K_{mj}^{(j)} \equiv \exp(\beta\mu_{mj}^{(j)})$ is the activity for a pure component monomer. In the thermodynamic limit $N^{(j)} \to \infty$ for the pure component material species $j = 1,2,3,\ldots$), various extensive quantities have well-defined densities per lattice site (see Refs. 10-14, and 42 for details). Thus, we define the adimensional pressure $z_0^{(j)} \equiv \omega^{(j)}$ for each of the pure components by the limiting value of the ratio $(1/N^{(j)})\ln Z_{N^{(j)}}^{(j)}$ as $N^{(j)} \to \infty$. We can follow the above derivation of Eqs. (10), and (11) to conclude that in equilibrium, we must have similar equations valid for each of the pure components:

$$(\partial\omega^{(j)}/\partial\phi_{mj}^{(j)}) = 0, \ (\partial\omega^{(j)}/\partial\phi_{0j}^{(j)}) = 0, \qquad (13)$$

$$(\partial s^{(j)}/\partial\phi_{mj}^{(j)}) = -\beta\mu_{mj}^{(j)}, \ (\partial s^{(j)}/\partial\phi_{0j}^{(j)}) = -\beta\varepsilon_{0j}. \qquad (14)$$

Here, $s^{(j)}$ is the entropy of the pure component per site.

Multiplying the thermodynamic densities by the size of a finite but very large lattice, we obtain extensive quantities for the lattice. One such quantity is the entropy *S*. According to the conventional definition of the entropy of mixing at constant *T* and *P*, we have

$$\Delta_M S \equiv S(T,P,\{N_{mj}\}|q,v_0,\{\varepsilon_{ij}\},\{M_j\}) - \sum_{j\geq 1} S^{(j)}(T,P,N_{mj}|q^{(j)},\varepsilon_{0j},v_0^{(j)},M_j). \qquad (15)$$

Note that the vacuum is *not* included in Eq. (15). This makes our definition different from that in Eq. (6). Replacing the entropy *S* by any other thermodynamic quantity *Q* in Eq. (15) gives us the mixing function $\Delta_M Q$. We also note that we have expressed *S* as a function of *T*, and *P* at the expense of the set $\{N_{ij}\}$, and *N*. In the process, it becomes a function of the relevant energy parameters, as shown explicitly in Eq. (15).

# IV. Some Cautionary Remarks
## (i) Lattice Properties

The values of the entropy $S$ and other thermodynamic potentials $E$, $F$, $G$, and $\tilde{P}$ depend strongly on the nature of the lattice. In particular, the coordination number $q$ and the connectivity of the lattice strongly affect their values. On the other hand, the cell volume $v_0$ is a special parameter, as it appears not in the partition function explicitly, but through the definition of what is meant by the pressure $P$, and can be easily absorbed in $\tilde{P}$. In this case, $N$ will play the role of the "volume" corresponding to $\tilde{P}$. The value of $v_0$ is not very crucial, as long as it remain *constant* so that $P$ and $\tilde{P}$ differ in scale only, and the equality of one implies the equality of the other in the system everywhere, as discussed in the previous section. It should be noted that several phenomenological theories treat the cell volume as an adjustable parameter, so that it depends on the thermodynamic state. We argue below that this is not allowed in a theory based on first principles.

The fixed and thermodynamic-state-independent nature of the lattice plays a very important role in the microscopic description of the lattice model, the importance of which has not been understood well. As a consequence, $q$ and $v_0$ must be independent of $T$, $P$, the composition, etc. To see it most easily, we recall that $V \equiv Nv_0$ must be an independent variable. Thus, it must be independent of the number of particles and, hence, the composition. Since $N$ is taken to be an independent variable in all lattice models, this implies that $v_0$ *cannot* be composition-dependent. In addition, for a composition-dependent $v_0$, the equality of $P$ and $\tilde{P}$ cannot be simultaneously maintained at coexistence between phases of different compositions. Some authors[35,40,41] have not appreciated this point and have allowed for composition-dependent $v_0$, which must be avoided at all cost. A composition-dependent $v_0$ gives rise to an incorrect description of polymer thermodynamics. The rigorous proof is given in the Appendix II. Thus, we must not use dressed parameters in place of bare parameters in the partition function; otherwise inconsistencies are bound to emerge.

We now turn our attention to the important issue of the choice of various lattices for the mixture and the pure components. We observe from Eq. (15) that for any of the quantities $\Delta_M Q$ like $\Delta_M S$ to vanish in the pure component limit in which all but one material species $j$ is present, we must either require that the mixture lattice change continuously so as to become identical with the pure $j$-component lattice, or that all lattices be *identical*. For the former case to hold will require the cell volume $v_0$ and $q$ of the mixture lattice to be composition-dependent so that they become $v_0^{(j)}$ and $q_0^{(j)}$, respectively, as the mixture turns into the pure *j*-component. This is not allowed in a microscopic theory as said above. As a consequence, *all lattices must be identical*. The requirement that $\Delta_M Q$ vanish in the pure component limit also requires that the DP's $M_j$ in the mixture and in the pure component be identical. Thus, we *cannot* allow for the DP $M_j^{(j)}$ in the pure component to be different from $M_j$. Unfortunately, many workers have failed to appreciate the importance of the above observations and have allowed different cell volumes $v_0$ and $v_0^{(j)}$. In addition, some authors[35,40,41] allow the same polymer to occupy *different* number of sites $M_j$ and $M_j^{(j)}$ on the mixture and the *j*-th pure component lattices, respectively, so that they are related: $M_j v_0 = M^{(j)} v_0^{(j)}$. Their theory goes further and takes $v_0$ to be a weighted average of $v_0^{(j)}$ to make $v_0$ composition-dependent. Accordingly, the volume of a void changes with composition, which is very hard to justify, since voids have no intrinsic size of any kind.

**(ii)     System Parameters**

The derivation of Eqs. (10), and (11) clearly shows why $\varepsilon_{ij}$ and $\mu_{mi}$ *cannot* be functions of the monomer, and contact densities in Eq. (7). The fundamental assumption of thermodynamics is that the Legendre transform in Eq. (9) from the densities $\{\phi_{mi}\}$ and $\{\phi_{ij}\}$, on which $s$ depends, to the fields $\beta, \{\varepsilon_{ij}\}$ and $\{\mu_{mi}\}$, on which $\omega$ depends, is such that $\omega$ is no longer a function of the densities. This basic requirement that all the parameters in the model be independent of the thermodynamic state of the system has not been fully appreciated by some authors. For example, many authors incorrectly allow the bare model parameter $\varepsilon_{ij}$ to have a composition-dependence. This composition-dependence, for example, has been ascribed by fitting the predictions of some theoretical calculation like the lattice-fluid theory to the experiments.[44]

Of course, it is possible to take some or all the parameters in the model like the cell volumes, coordination numbers, etc. to be different in the mixture and the pure components at the expense of relaxing $\Delta_M Q = 0$ in the pure component limit, as long as these quantities remain constant and remain independent of the thermodynamic state of the mixture. This is because the thermodynamics of the mixture cannot be affected by how pure components behave.

Consider, for example, the issue of the *configurational entropy*, which represents the entropy of a single polymer with one of its end fixed at the origin of an otherwise empty lattice, assuming no interactions. Evidently, it depends on the lattice coordination number and the degrees of polymerization. Let $\ln f_j$ and $\ln f^{(j)}$ denote the configurational entropy per $j-$species polymer on the mixture and the pure component lattices, respectively. The *j*-th species configurational entropy contribution $\Delta_M S_{\text{conf}}^{(j)}$ to the entropy of mixing is, therefore,

$$\Delta_M S_{\text{conf}}^{(j)} = N_j \ln(f_j / f^{(j)}).$$

Unless the lattices used for the mixture and the pure components are the same, $f_j$ and $f^{(j)}$ will be different and $\Delta_M S_{\text{conf}}^{(j)}$, which no longer vanishes, should be included in $\Delta_M S$. Furthermore, its value will depend on the lattices that are used and so will the value of $\Delta_M S$. The total configurational entropy contribution only adds a term *linear* in $N_j$ as long as $f_j$ and $f^{(j)}$ remain *constant*. In this case, its contribution to the second derivative with respect to $N_j$ vanishes. Thus, it will not affect the phase boundary, and need not be included in $\Delta_M S$. If $f_j$ and/or $f^{(j)}$ change with composition, as happens in the lattice fluid theory[35,40,41] due to a difference between $M_j$ and $M_j^{(j)}$, then the second derivative of $\Delta_M S_{\text{conf}} = \sum_{j\geq 1}\Delta_M S_{\text{conf}}^{(j)}$ with respect to composition will *not* vanish. In this case, $\Delta_M S_{\text{conf}}^{(j)}$ must be included in $\Delta_M S$ and other mixing functions. However, we have already argued that this aspect of the lattice fluid theory is incorrect. We, therefore, assume all these parameters in the model to be the same in the mixture and in the pure components. In addition, we do not impose any *ad hoc* mixing rules in our theory. They could be deduced from the fit of our theory with experiments.

## V. Recursive Lattice Theory

The entropy and other thermodynamic functions of a multi-component compressible system have been calculated by Ryu and Gujrati[42] and by Gujrati[12] in a scheme that goes beyond the RMA, but contains it as a special limit; see (iii) below. It is a general theory and treats monodisperse and polydisperse species of any architecture. The limitations and strengths of the approach have been discussed elsewhere.[10,11,45-49]

**(i) Mixture**

We quote below the results for a general multi-component compressible mixture. We should remark that the results in Refs. 42 and 12 are given for extensive thermodynamic densities per lattice site. Thus, there is a change in notation here from our earlier works where uppercase symbols *S*, *E*, etc. refer not to the total but to the density values. Here, we give results for total extensive functions. We consider only monodisperse species. The total entropy *S* is given by

$$S(\{N_j\},\{N_{ij}\}|q,\{M_j\}) = \sum_{j\geq 0} N_j \ln(f_j/\phi_{nj}) + N_u \ln(2\phi_u/q) + \sum_{i\leq j} N_{ij} \ln(\phi_{ij}^0/\phi_{ij}), \quad (16)$$

where the first sum is over all species $j \geq 0$, and the last sum is over distinct pairs of species. The quantity $f_j$ denotes the *embedding constant* of a *j*-species molecule (polymer, solvent or void). It is equal to the number of distinct ways a polymer can be put on an otherwise empty Bethe lattice of coordination number $q$, such that an end point is located at the origin of the lattice. For species occupying a single site of the lattice, $f = 1$. For a linear polymer containing *b* bonds and the two indistinguishable end-points, $f = (1/2)qr^{b-1}$; here $r \equiv q-1$. The embedding constant depends on the architecture of the species and is easily calculated. The remaining quantities are the number (n) density $\phi_{nj} \equiv N_j/N$, the total number of bonds $B \equiv \sum N_j(M_j - 1)$, the total bond density $\phi \equiv B/N$, the number of lattice bonds left uncovered (u) by polymers $N_u = qN/2 - B$, the density of lattice bonds uncovered by polymers $\phi_u = q/2 - \phi$, and the contact density is defined earlier.

The monomer density can be written as $\phi_{mi} \equiv M_i \phi_{ni}$ and the bond density as $\phi_i \equiv (M_i - 1)\phi_{ni}$ for the *i*-th species. Then the density of lattice bonds attached to *j*-th species monomers is twice the density $\phi_{ju} = q\phi_{mj}/2 - \phi_j \equiv q_j\phi_{mj}/2$, where we have introduced a new quantity $q_j \equiv q - 2\nu_j$, and $\nu_j \equiv 1 - 1/M_j$. The contact densities with superscript 0 are their athermal values, when all interactions vanish, and are given by

$$\phi_{jj}^0 = \phi_{ju}^2 / \phi_u, \quad \phi_{ij}^0 = 2\phi_{iu}\phi_{ju}/\phi_u \quad (i \neq j). \tag{17}$$

The contact density $\phi_{ij}$, $i \neq j$, is given by

$$\phi_{ij} = 2w_{ij}\sqrt{\phi_{ii}\phi_{jj}}. \tag{18}$$

For the athermal case, we have $\phi_{ij}^{(0)} = 2\sqrt{\phi_{ii}^{(0)}\phi_{jj}^{(0)}}$.

The first sum in Eq. (16) contains the contribution from *all* species including voids and is *always* non-negative. The middle term gives the contribution of the chemical bonding, is negative and vanishes in the absence of chemical bonding. The last sum is the contribution from contact densities, is negative and vanishes in the athermal limit.

The chemical potential $\mu_{mj}$ of the $j$-th species monomer is given by

$$\begin{aligned}\beta\mu_{mj} &= (1/M_j)\ln(\phi_{nj}/f_j) - \ln\phi_0 + \nu_j \ln(2\phi_u/q) \\ &+ (q_j/2)\ln(\phi_{jj}/\phi_{jj}^0) - (q/2)\ln(\phi_{00}/\phi_{00}^0)\end{aligned} \tag{19}$$

The adimensional pressure is given by

$$z_0 \equiv \beta P v_0 = -\ln\phi_0 + (q/2)\ln(2\phi_u/q) + (q/2)\ln(\phi_{00}^0/\phi_{00}). \tag{20}$$

It should be noted again that the cell volume $v_0$ appears only in the pressure equation (20), and nowhere else. Thus, as expected, the actual value of the cell volume only affects the volume $V$, and the pressure $P$ individually but not their product $\tilde{P} \equiv Pv_0$, which has the dimension of work or energy and remains invariant.

**(ii) Pure Components**

As discussed in the previous section, we assume that the lattice required for the pure component is the *same* as the one used for the mixture. Thus, the cell volume for the pure component lattice is also taken to be $v_0$. Introducing $N^{(j)}$ for the number of sites in the pure component lattice, and other quantities and densities appropriately, we have

$$\begin{aligned}S^{(j)}(N_j, N_0^{(j)}|q, v_0, M_j) &= -N_0^{(j)}\ln\phi_0^{(j)} + N_j\ln(f_j/\phi_n^{(j)}) + N_u^{(j)}\ln(2\phi_u^{(j)}/q) \\ &+ N_{00}^{(j)}\ln(\phi_{00}^{(j)0}/\phi_{00}^{(j)}) + N_{01}^{(j)}\ln(\phi_{01}^{(j)0}/\phi_{01}^{(j)}) + N_{11}^{(j)}\ln(\phi_{11}^{(j)0}/\phi_{11}^{(j)}),\end{aligned} \tag{21a}$$

$$\begin{aligned}\beta\mu_{mj}^{(j)} &= (1/M_j)\ln(\phi_n^{(j)}/f_j) - \ln\phi_0^{(j)} + \nu_j\ln(2\phi_u^{(j)}/q) \\ &+ (q_j/2)\ln(\phi_{jj}^{(j)}/\phi_{jj}^{(j)0}) - (q/2)\ln(\phi_{00}^{(j)}/\phi_{00}^{(j)0}),\end{aligned} \tag{21b}$$

$$\beta Pv_0 = -\ln\phi_0^{(j)} + (q/2)\ln(2\phi_u^{(j)}/q) + (q/2)\ln(\phi_{00}^{(j)0}/\phi_{00}^{(j)}), \tag{21c}$$

from Eqs. (16,19,20). The superscript $(j)$ denotes the $j$-th pure component. The numbers $N^{(j)}$ and $N_0^{(j)}$ are determined by requiring that the pure component has the same temperature and pressure as the mixture. We must choose a particular value of $N$ for this purpose. Otherwise, what one determines are the densities $\phi_0^{(j)}$.

### (iii) RMA Limit

The RMA limit[12] in the compressible lattice model is *equivalent* to taking the simultaneous limits $q \to \infty$, $\beta \to 0$ and $P \to \infty$, such that $\chi_{ij}$ and $z_0$ are held *fixed* and *finite*. In the limit, the contact densities

$$\phi_{ij} \to \phi_{ij}^{(0)} \tag{22a}$$

for the mixture, and

$$\phi_{0j}^{(j)} \to \phi_{0j}^{(j)0}, \quad \phi_{jj}^{(j)} \to \phi_{jj}^{(j)0} \tag{22b}$$

for the pure components. We also find that in this limit quantities like $q \ln(2\phi_u/q)$ have a simple limiting behavior:

$$q \ln(2\phi_u/q) \to -2\phi, \tag{22c}$$

and similar relations for the pure components. Using these limiting behavior, we find that the equation of state for the mixture and the pure components in the RMA limit reduces to the forms given below,

$$\begin{aligned} z_{0,\text{RMA}} &= -\ln \phi_0 - \phi - \sum_{j>0} \chi_{0j} \phi_{mj} + \sum_{j>i\geq 0} \chi_{ij} \phi_{mi} \phi_{mj}, \\ z_{0,\text{RMA}}^{(j)} &= -\ln \phi_0^{(j)} - \phi^{(j)} - \chi_{0j} (\phi_{mj}^{(j)})^2. \end{aligned} \tag{23}$$

The first equation is for the mixture, and the second equation is for the pure component. It is easy to check that our RMA equation of state for the mixture is *different* from that due to Sanchez and Lacombe,[35,40,41] even though they are the same for the pure component, as was noted earlier in Ref. 12. The most important difference is the absence of $j = 0$ term in the last sum in the first equation in Eq. (23). There is also some difference in the first sum in the same equation. The first equation is easily derived from the following equation due to Ryu[50] relating $\phi_{00}$ with other densities. We will only give the results for a binary blend. The extension to the general case is trivial and is given in the Appendix III. From (A.3.4) in the Appendix III,

$$\phi_{00} = \overline{D}_0^2 / D, \tag{24a}$$

where

$$D = \phi_u + \overline{w}_{01}\phi_{01} + \overline{w}_{02}\phi_{02} + \overline{w}_{12}\phi_{12}, \quad \overline{D}_0 = \phi_{0u} + (\overline{w}_{01}\phi_{01} + \overline{w}_{02}\phi_{02})/2, \tag{24b}$$

and where $\overline{w}_{ij} = 1/w_{ij} - 1 \to \varepsilon_{ij} \to 0$ in the RMA limit. Therefore, we keep the first-order terms in $\varepsilon_{ij}$ in the ratio $\phi_{00}^0/\phi_{00}$, with $\phi_{00}^0$ given in Eq. (17) to obtain Eq. (24a).

In contrast, the two customary forms for the extension of the F-H theory to a compressible system are not correct, as discussed in Sec. III. One of them replaces the ideal entropy of mixing by the use of volume fractions. This form cannot account for volume of mixing. The other extension is obtained by the mixing process used in deriving Eq. (6). This extension has unphysical behavior as we have discussed. In particular, it gives rise to an unphysical effective dressed chi $\chi_{\text{eff,LF}}$; see Eq. (39) and subsequent discussion thereafter. Our theory provides us with a proper extension in the RMA limit for a compressible system, and is given in Eqs. (27), (28), (35), and (49). This extension gives rise to a sensible $\chi_{\text{eff,RMA}}$ in Eq. (40).

## VI. Volume of Mixing

From now onward, we will mostly study a binary mixture extensively, even though the results are also given for a general mixture. To simplify the notation for a blend, we will use one and two primes to denote quantities pertaining to the two pure components, and no prime for quantities pertaining to the mixture. We will also use $y \equiv y_2$ to represent the composition. We need to make a distinction between the following two cases, as the behavior is very different in each case.

(1) *Symmetric blend* ($M' = M''$ and $w_{01} = w_{02}$).
(2) *Asymmetric blend* ($M' \neq M''$ and/or $w_{01} \neq w_{02}$).

It should be stressed that our definition of a symmetric blend differs from its conventional definition in that an asymmetry in the interaction alone also qualifies the blend to be asymmetric. This distinction must be kept in mind.

The volume of mixing for a binary mixture $\Delta_M V = V - V' - V''$, which also represents the change in the free volume $\Delta_M V_0 = V_0 - V_0' - V_0''$, has been calculated in Eq. (20) in Ref. 12, where it is denoted simply by $\Delta V$. Normalizing this by the total number of monomers $N_m$, noting that $N_m / V = (1-\phi_0)/v_0$, we find that the volume of mixing per monomer $\Delta_M v_m \equiv \Delta_M V / N_m \equiv \Delta$, and the volume of mixing per unit volume $\Delta v \equiv (1-\phi_0)\Delta_M v_m$, and where $\Delta$ is given in Eq. (21) in Ref. 12:

$$\Delta_M v_m = v_0 \left( 1/(1-\phi_0) - y_1/(1-\phi_0') - y_2/(1-\phi_0'') \right). \tag{25a}$$

The void fractions with primes refer to the pure components. For a multi-component mixture, Eq. (25a) generalizes to

$$\Delta_M v_m = v_0 \left( 1/(1-\phi_0) - \sum_{j \geq 1} y_j /(1-\phi_0^{(j)}) \right). \tag{25b}$$

Here,

$$y_j \equiv M_j N_j / N_m$$

denote the monomer fractions.

We now consider an athermal binary mixture. A simple reasoning that is found in Ref. 12 shows that $\phi_0 = \phi_0' = \phi_0''$ for a symmetric mixture, so that $\Delta_M v_m \equiv 0$. On the other hand, $\phi_0$ lies between $\phi_0'$, and $\phi_0''$ for an asymmetric mixture, so that $\Delta_M v_m < 0$. As we increase $q$, $\Delta_M v_m$ increases and approaches an asymptotic negative value, as we see from the behavior of $\Delta v$, see Fig. 3a, at $y = 1/2$ for an asymmetric blend with Dp's 100, and 10. The reduced pressure is $z_0 = 0.2$. The effect of interactions on $\Delta v$, as shown in Fig. 3b, is expected. For repulsive (attractive) interactions between the species, the mixture volume expands (contracts) relative to its athermal value. For a symmetric blend, therefore, $\Delta v$ is positive for $w \equiv w_{12} < 1$ and negative for $w > 1$, as shown in Fig. 3b for equal DP 100. It remains negative for $w > 1$ and can become positive for $w < 1$ for an asymmetric blend. The effect of pressure in this case is interesting. For the attractive case, $\Delta v$ becomes more and more negative as we increase the pressure. However, for the repulsive case, $\Delta v$ can become positive as we increase the pressure, as shown in Fig. 3c. At an intermediate pressure $z_0 \cong 0.177$, there appears a zero in $\Delta v$ at a certain composition, as shown in Fig. 3d. We refer the reader to Ref. 12 for details.

According to the modified regular solution theory,[4-6] which allows for non-zero volume of mixing,

$$\Delta_M v_m = \alpha(P,T) y_1 y_2, \quad \bar{v}_1 = \bar{v}_1^0 + \alpha(P,T) y_2^2, \quad \bar{v}_2 = \bar{v}_2^0 + \alpha(P,T) y_1^2, \tag{26}$$

where $\bar{v}_i$, and $\bar{v}_i^0$ denote the $i$-th species partial monomer volumes in the mixture and in the $i$-th species pure component, respectively, and $\alpha$ is a function independent of composition. It is easily seen that with a composition-independent $\alpha$, the last two equations for the partial monomer volumes can be easily derived from the first equation. Accordingly, $\Delta_M v$ is expected to have either a maximum or a minimum at $y = 1/2$. Since $\phi_0$ remains non-zero in a compressible blend, this means that there cannot be a zero in $\Delta v$ as a function of $y$. This is evidently not true, as we see From Fig. 3d. Thus, $\alpha$ is, in general, not composition-independent, and the three equations in Eq, (26) are mutually inconsistent. Indeed, we have shown elsewhere[51] that the behavior of partial monomer volumes in Eq. (26) cannot be justified in general. Thus, the modified regular solution theory is *not* valid in general, except in some special and limited cases. Even for a symmetric blend, for which the volume of mixing must be symmetric in $y$, the quantity $\alpha$ turns out to be weakly composition-dependent.[51] For asymmetric blends, it becomes strongly composition-dependence so as to accommodate the volume of mixing that can change its sign, as in Fig. 3d.

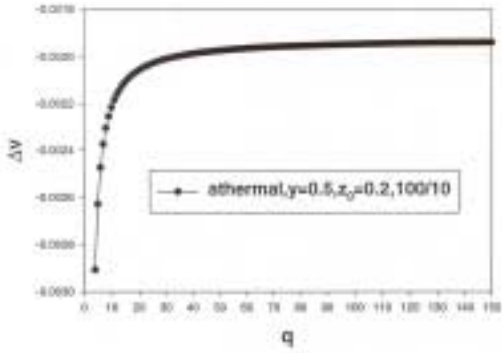

(a)

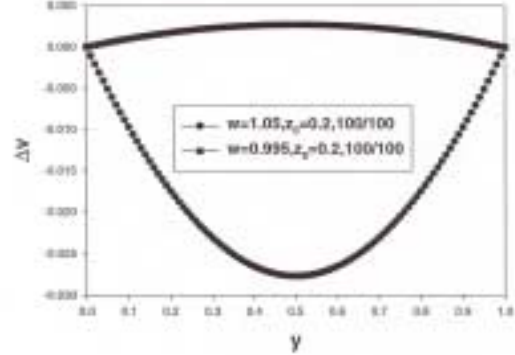

(b)

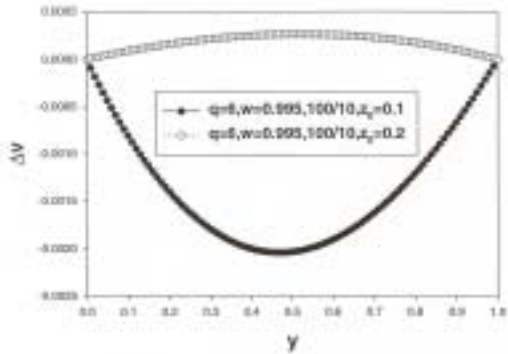

(c)

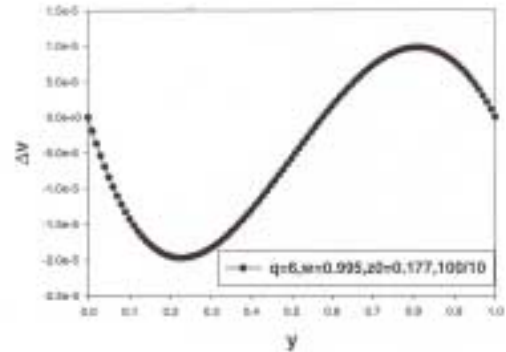

(d)

Fig. 3 (a) Effect of varying $q$ on $\Delta v$ at $y \equiv y_2 = 1/2$ in an athermal asymmetric blend. (b) $\Delta v$ as a function of $y$ for an interacting symmetric blend at fixed adimensional pressure; $q = 6$. (c,d) The effect of pressure on $\Delta v$ for asymmetric repulsive blend.

## VII. Entropy of Mixing

### (i) Total Entropy of Mixing

We begin by focusing on a binary mixture and consider various parts of the total entropy of mixing before considering a multi-component mixture. However, for the latter, we only give the results for the entropy of mixing per monomer or particle. The first two terms in Eq. (16) corresponds to the entropy when there are no interactions. Thus, they represent the entropy in the athermal state. Thus, to investigate these two terms, we can assume that there are no interactions in the system. The effects of interactions appear in the third term in that equation. As interactions always reduce the entropy, this term must always be negative. This follows immediately from the lemma proved in Ref. 52.

**(a) COM Contribution.** The *center-of-mass* (COM) contribution from $j=1,2$ in the first sum in Eq. (16) is given by

$$\Delta_M S_{COM} = N_1 \ln(\phi'_{n1}/\phi_{n1}) + N_2 \ln(\phi''_{n2}/\phi_{n2}) = -N_1 \ln v' - N_2 \ln v'', \tag{27}$$

which is identical in *form* to Eq. (2), *except* that $\Delta_M V$ need *not* vanish. This contribution is merely due to the placement of the COM's on lattice sites.

**(b) Free Volume Contribution.** The void contribution ($j=0$) from the first sum in Eq. (16), which we will term the *free-volume* (fv) contribution, is given by

$$\Delta_M S_{fv} = -N'_0[\ln(V'/V) - \ln(V'_0/V_0)] - N''_0[\ln(V''/V) - \ln(V''_0/V_0)] \\ - \Delta_M N_0 \ln(V_0/V), \tag{28}$$

where $\Delta_M N_0 \equiv N_0 - N'_0 - N''_0$. Its form is very *different* from that in Eq. (3) due to Hildebrand.[6,25] Introducing the average monomer volume $v_m \equiv V/N_m = v_0/\phi_m$, $v'_m \equiv V'/N_{m1} = v_0/\phi'_m$, and $v''_m \equiv V''/N_{m2} = v_0/\phi''_m$ for the mixture and the pure components, and using monomer fractions $y_1, y_2$, we have

$$v_m = y_1 v'_m + y_2 v''_m, \quad \phi_0 v_m = y_1 \phi'_0 v'_m + y_2 \phi''_0 v''_m, \tag{29}$$

when $\Delta_M V = 0$. For Eq. (29) to be valid for all values of $y_1$ or $y_2$, we must have

$$\phi_0 = \phi'_0 = \phi''_0. \tag{30}$$

Accordingly, $\Delta_M S_{fv}$ vanishes for isometric mixing. As we discussed above,[12] Eq. (30) is valid only for athermal symmetric blends.

**(c) Unbonded Bond Contribution.** The contribution from lattice bonds that are chemically unbonded (unb), i.e., left uncovered by polymers is given by

$$\Delta_M S_{unb} \equiv N'_u \ln(\phi_u/\phi'_u) + N''_u \ln(\phi_u/\phi''_u) + (q/2)\Delta_M N \ln(2\phi_u/q), \tag{31}$$

where $\Delta_M N = \Delta_M V/v_0$, $N'_u = qN'/2 - B'$, $\phi'_u = N'_u/N' = q/2 - \phi'$, etc. The mixing process does not change the number of bonds; hence, $B \equiv B' + B''$.

The contribution in Eq. (31) vanishes if all material species are monomeric. It also *vanishes* for isometric mixing for which, because of Eq. (30), all three uncovered bond densities are equal. Only $\Delta_M S_{COM}$ in Eq. (27) survives in this case. For an interacting or asymmetric blend, $\Delta_M V$ and, therefore, the other two contributions need not vanish. *Non-zero contributions in Eqs.* (28), *and* (31) *are a consequence of non-isometric mixing.*

**(d)     Contact Contribution.** The last contribution to be denoted by a subscript "cont", from various contact densities can not be reduced to any convenient-looking form and are always present because of interactions and provide corrections to the regular solution theory ($q$ finite). There are six terms in the last sum in Eq. (16). Each of the two pure components contributes three terms, as shown in Eq. (21a). Thus, there are twelve terms in total, which we do not write down here explicitly

**(ii)     Athermal Limit**

It should be obvious from the derivation leading to Eq. (29) that it must be valid in all viable theories, not just our theory. A viable theory must yield an equation of state relating $P$ and $T$ to the density, which in turn must be related to the void density. In addition, the equation must also depend on parameters like $M_j$, $q$, etc. It is evident that, since $P$ and $T$ are the same for all pure components, and that interactions are not present (athermal condition), the free volume density *cannot* be the same for the two pure components unless the polymer DP's are the same. Hence, athermal mixing *cannot* be isometric for *different* DP's, except when the system is incompressible. Therefore, we are forced to conclude that if the DP's are different and the system is compressible, then the regular solution theory cannot be a suitable theory for the model in *any* limit, except when the system is incompressible.

Even in the athermal limit, the entropy of mixing is not of the conventional type, unless we neglect the contributions in Eqs. (28), and (31). For a symmetric blend, for which $\Delta_M V = 0$, we retrieve the ideal form of the entropy of mixing. At least in our theory, *the ideal entropy of mixing is valid for an athermal mixture of symmetric particles, no matter what their size.* The result is inconsistent with Eq. (3), which contains free volumes and not the volumes of the three systems. We begin to see deviation from ideality when $\Delta_M V \neq 0$. The first change is in Eq. (27) in which the volume ratios do *not* add to one. In addition, there are other two non-zero contributions that appear in Eqs. (28), and (31). The entropy of mixing contains both the volumes and the free volumes.

**(iii)    RMA Limit**

In the RMA limit, the last two contributions $\Delta_M S_{unb}$, and $\Delta_M S_{cont}$ in the entropy of mixing vanish. This is easily seen from the limiting forms of various quantities in Eq. (22). For example, it is easily concluded that $\Delta_M S_{unb}$ decreases as $q$ increases and vanishes as $q \to \infty$; see, for example, Fig. 4(a). Thus, we are left with the first two contributions in Eqs. (27), and (28), neither of these contain $q$ explicitly. Hence, it is not surprising that the entropy of mixing in Eqs. (32a, b) can be put in a form suitable for continuum description. However, it is also clear that even in this limit, our entropy of mixing is *different* from that in Eq. (4).

### (iv) Contributions per monomer

Let us evaluate the entropy of mixing $\Delta_M s_m$ *per particle* by dividing the above contributions by the total number of particles $N_m$. We give the results valid for a multicomponent mixture. There are four parts to this entropy, which are as follows.

$$\Delta_M s_{m,\text{COM}} = -\sum_{j\geq 1}(y_j/M_j)\ln(\phi_{mj}/\phi_m^{(j)}) = -\sum_{j\geq 1}(y_j/M_j)\ln(V^{(j)}/V). \quad (32a)$$

The free-volume contribution is

$$\Delta_M s_{m,\text{fv}} = -(\phi_0/\phi_m)\ln\phi_0 + \sum_{j\geq 1} y_j(\phi_0^{(j)}/\phi_m^{(j)})\ln\phi_0^{(j)}$$
$$= -[(v_m - v_0)/v_0]\ln(V_0/V) + \sum_{j\geq 1} y_j[(v_m^{(j)} - v_0)/v_0]\ln(V_0^{(j)}/V^{(j)}), \quad (32b)$$

where the monomer densities $\phi_m = 1 - \phi_0$ and $\phi_m^{(j)} = 1 - \phi_0^{(j)}$, and $v_m^{(j)} \equiv V^{(j)}/N_{mj}$ is the average volume per monomer for pure components. The unbonded lattice-bond contribution is

$$\Delta_M s_{m,\text{unb}} = (\phi_u/\phi_m)\ln(2\phi_u/q) - \sum_{j\geq 1} y_j(\phi_u^{(j)}/\phi_m^{(j)})\ln(2\phi_u^{(j)}/q),$$
$$= (q/2v_0)[v_m(1-\xi)\ln(1-\xi) - \sum_{j\geq 1} y_j v_m^{(j)}(1-\xi^{(j)})\ln(1-\xi^{(j)})], \quad (32c)$$

where we have introduced new quantities $\xi$ and $\xi^{(j)}$ for the mixture and the pure components, respectively, as follows.

$$\xi = 2\sum_{i\geq 1}(v_i y_i)v_0/v_m q, \quad \xi^{(j)} = 2v_j v_0/v_m^{(j)} q.$$

These quantities depend on the coordination number of the lattice; hence, they *cannot* have a continuum analog. The contribution in Eq. (32c), therefore, cannot possess a continuum form. This is sufficient to show that there is no possibility of casting lattice results in a form suitable for a continuum picture.

The last contribution is from contacts and is given by

$$\Delta_M s_{m,\text{cont}} = \sum_{0\leq i\leq j}(\phi_{ij}/\phi_m)\ln(\phi_{ij}^0/\phi_{ij}) - \sum_{j\geq 1} y_j(\phi_{0j}^{(j)}/\phi_m^{(j)})\ln(\phi_{0j}^{(j)0}/\phi_{0j}^{(j)}). \quad (32d)$$

The contact densities also cannot be put in a form suitable for a continuum version, since these densities are related to bonds pertaining to certain kinds of nearest-neighbor contacts and must necessarily depend on the coordination number. We have already shown, see Fig. 3 in Ref. 47, that the contact densities are not simply proportional to $q$. Moreover, the contact densities are complicated functions of $\varepsilon_{ij}$; hence the above contribution will also change with $q$, as we will show in Fig. 4 below.

## VIII. Energy of Mixing
### (i) Energy of mixing

We proceed with a general mixture and consider a binary mixture whenever necessary. The energy of mixing is given by

$$\Delta_M E = \sum_{j\geq 1}\varepsilon_{0j}(N_{0j} - N_{0j}^{(j)}) + \sum_{1\leq i<j}\varepsilon_{ij}N_{ij}. \quad (33)$$

If we add $\beta P\Delta_M V$ to $\Delta_M \beta E$, we obtain the adimensional enthalpy of mixing. We will not give the explicit expression for $\Delta_M \beta H$. The second sum in Eq. (33) is purely a mixture property. The first sum, however, depends on the pure components also. Thus, in general, the energy of mixing *cannot* be taken as a measure of the interactions in the mixture alone. Only in the incompressible limit, where the first sum vanishes, will the energy of mixing be a measure only of the interactions in the mixture. Even here, the second sum in Eq. (33) depends on the thermodynamic state of the mixture, which determines the values of the contacts $N_{ij}$. Thus, even for an incompressible system, extracting the interaction parameters $\varepsilon_{ij}$ is not feasible, unless we know exactly $N_{ij}$. In an exact theory, we can invert the above relationship in Eq. (33) to express $\varepsilon_{ij}$ in terms of the energy of mixing and other state variable. Consider, for example, a binary blend and assume that we know $\varepsilon_{0j}$ related to its two pure components. We can express $\varepsilon_{12}$ by inverting Eq. (33). Unfortunately, we do not have an exact theory at present even for a binary mixture. Hence, we are limited in our ability to find $\varepsilon_{12}$ precisely, which is an important model parameter as discussed in the Introduction. Use of an approximate theory leads to an *effective* or a *dressed* chi $\chi_{\text{eff}}$ as an estimator of $\varepsilon_{12}$. The estimator depends on the thermodynamic state of the system, and is introduced below. The aim is to find a suitable $\chi_{\text{eff}}$ that is weakly dependent on the thermodynamic state and, in particular, on the compressibility.

In the RMA limit, the average contact densities have a simple form:
$$N_{ii} = (q/2)N_{\text{m}i}\phi_{\text{m}i}, \qquad N_{ij} = qN_{\text{m}i}\phi_{\text{m}j}, \quad i \neq j,$$
see Eqs. (22a), and (17). In this limit, $\chi_{ij}$ are kept fixed and finite. Thus, we find that
$$\Delta_M \beta E_{\text{RMA}} = \sum_{1 \leq i < j} \chi_{ij} N_{\text{m}i}\phi_{\text{m}j} + \sum_{j \geq 1} \chi_{0j} N_{\text{m}j}(\phi_0 - \phi_0^{(j)}). \tag{34}$$
The interesting aspect of Eq. (34) is that it is in a form suitable for continuum interpretation. For a binary blend, Eq. (34) reduces to
$$\Delta_M \beta E_{\text{RMA}} = \chi_{12} N_{\text{m}1}\phi_{\text{m}2} + \chi_{01} N_{\text{m}1}(\phi_0 - \phi_0') + \chi_{02} N_{\text{m}2}(\phi_0 - \phi_0''). \tag{35}$$
In the incompressible limit, the last two contributions in Eq. (35) vanish, and the remaining contribution requires only knowing the composition of the blend. Thus, we can extract the value of the mixture parameter $\chi_{12}$. However, as soon as we leave the incompressibility limit, this is no longer true. The composition of the mixture *alone* is not sufficient to determine the mixture contribution to the energy of mixing.

The energy of mixing per monomer from Eq. (33) is
$$\Delta_M e_m = \sum_{1 \leq j} \varepsilon_{0j}\left[\left(\phi_{0j}/(1-\phi_0) - y_j \phi_{0j}^{(j)}/(1-\phi_0^{(j)})\right)\right] + \sum_{1 \leq i \leq j} \varepsilon_{ij}\phi_{ij}/(1-\phi_0). \tag{36}$$
In the RMA, we find from Eq. (35) that the adimensional energy of mixing is given by
$$\Delta_M \beta e_{m,\text{RMA}} = \sum_{j \geq 1} \chi_{0j}(\phi_0 - \phi_0^{(j)})y_j + \sum_{1 \leq i \leq j} \chi_{ij} y_i \phi_{\text{m}j}. \tag{37}$$

**(ii) Effective chi**

Even though the energy of mixing is not a measure of the interaction between the mixing components, except in the incompressible limit, it may still play a useful role as a measure. We adopt this viewpoint and define an effective chi for a binary mixture that could play such a role. This effective chi is defined as follows:

$$\chi_{\text{eff}} \equiv (\beta \Delta_M E / N) / \phi_{m1} \phi_{m2}. \tag{38}$$

Of course, we can define an effective chi, not by Eq. (38), but by some suitable derivative of $\Delta_M E$ with respect to composition; see below. The above choice is just one of many convenient and useful choices. We will first consider the "energy of mixing" given by Sanchez and coworkers.[40,41] It is given by

$$\Delta_M \beta E_{LF} / N = \chi_{12} \phi_{m1} \phi_{m2} + (\chi_{01} \phi_{m1} + \chi_{02} \phi_{m2}) \phi_0.$$

(We will use the subscript LF to denote quantities in this theory.) This immediately yields for $\chi_{\text{eff}}$ the following expression:

$$\chi_{\text{eff,LF}} = \chi_{12} + (\chi_{01} / \phi_{m2} + \chi_{02} / \phi_{m1}) / \phi_0. \tag{39}$$

It has a *disturbing* feature that it diverges in the wings where the monomer densities vanish. This unphysical feature is a consequence of their special mixing process, as discussed in Sect. III(d).

The proper definition of $\Delta_M E$ in Eq. (33) that we utilize in Eq. (38) does *not* give rise to this spurious divergence in $\chi_{\text{eff}}$, because the difference between $N_0$ and $N_0^{(j)}$ approaches zero as we approach the pure *j*-state. In particular, in the RMA limit,

$$\chi_{\text{eff,RMA}} = \chi_{12} + \chi_{01} (\phi_0 - \phi_0') / \phi_{m1} + \chi_{02} (\phi_0 - \phi_0'') / \phi_{m2}. \tag{40}$$

We note from Eq. (38) that $\chi_{\text{eff}}$ is directly related to $\Delta_M e_m \equiv \Delta_M E / N(1-\phi_0)$: We have

$$\chi_{\text{eff}} \equiv (1-\phi_0) \beta \Delta_M e_m / \phi_{m1} \phi_{m2} = [\beta \Delta_M e_m / y_1 y_2]/(1-\phi_0). \tag{41}$$

We can also use the following derivative to introduce another dressed or effective chi for $\chi_{12}$:

$$\chi'_{\text{eff}} \equiv -[1/2(1-\phi_0)](\partial^2 \beta e_m / \partial y^2)_{T,N}. \tag{42}$$

The prefactor in Eq. (42) ensures that $\chi'_{\text{eff}} = \chi_{12}$ in the incompressible RMA limit.

Both $\chi_{\text{eff}}$, and $\chi'_{\text{eff}}$ reduce to $\chi_{12}$ only in the incompressible RMA limit but are different from it in general. In general, we find that both quantities are close to

$$\chi_{\text{NR}} \equiv [(q-2)/q] \chi_{12}, \tag{43}$$

for symmetric blends with small amount of free volume (as shown in Sect. X), because of the corrections due to polymer connectivity, which reduces the possible number of contacts between dissimilar species from $q$ to $(q-2)$. Here, we are neglecting the end-group effects, i.e. we are assuming the polymers to be very large in size.

From the above, it is evident that, while $\Delta_M E$ certainly represents the change in the interaction energy due to mixing, it does *not* truly represent the 1-2 interaction unless the mixture is incompressible. Thus, the effective chi's defined above also are not true estimator of the 1-2 interaction. The additional contribution due to compressibility gives rise to features that, in some cases, are counter-intuitive, as we see in the next subsection.

**(iii)   Cohesive Energy Density & Solubility Parameter.**

The energy per unit volume of the *j*-th pure component in terms of $e_{jj}$ is given by

$$E_j^{(j)}/V^{(j)} \equiv e_{jj}\phi_{jj}^{(j)}/v_0, \qquad (44)$$

where $E_j^{(j)} \equiv e_{jj}N_{jj}^{(j)}$. In the lattice model, there is no *kinetic energy* contribution. Thus, $E_j^{(j)}$ also represents the energy of vaporization, since the energy and the energy density in Eq. (44) of the (infinite volume) vapor state are identically zero. The energy of vaporization is an integrated quantity and contains the discontinuity in the energy across the liquid-vapor coexistence. The cohesive energy density $c_{jj}$ is given by

$$c_{jj} \equiv \delta_{jj}^2 \equiv -e_{jj}\phi_{jj}^{(j)}/v_0, \qquad (45)$$

where we have also introduced the solubility parameter $\delta_{jj}$. At this stage, Eq. (45) is merely a definition of $c_{jj}$ and $\delta_{jj}$. Being an integrated quantity, the cohesive energy density *cannot*, in general, be equal to $[-(\partial E_j^{(j)}/\partial V^{(j)})_{T,N_{mj}}]$, which is a differential quantity, except possibly in some special cases.[53]

In the RMA, the energy density of the pure component is given by

$$E_{j,\text{RMA}}^{(j)}/V^{(j)} = (-\chi_{0j}/\beta)v_0/(v_m^{(j)})^2, \qquad (46)$$

which remains valid in both lattice and continuum formulations. Thus,

$$c_{jj,\text{RMA}} \equiv \delta_{jj,\text{RMA}}^{(2)} = (\chi_{0j}/\beta)v_0/(v_m^{(j)})^2. \qquad (47)$$

Consider now a binary mixture. The last two terms in Eq. (40) are absent in the case of isometric mixing and $\chi_{\text{eff,RMA}} = \chi_{12}$, which is a constant independent of composition, DP's, pressure, etc. Moreover, the average monomer volumes are the same for the mixture and the pure components, which we denote by $\bar{v}_m$. If we use the well-known but much abused London-Berthelot conjecture $(-q\beta e_{12}) = 2\sqrt{\chi_{01}\chi_{02}}$, we find that

$$\chi_{\text{eff,RMA}} = (\sqrt{\chi_{01}} - \sqrt{\chi_{02}})^2 = \bar{v}_m^2 \beta(\delta_{11,\text{RMA}} - \delta_{22,\text{RMA}})^2/v_0. \qquad (48)$$

Since $\bar{v}_m$ depends on $P$, it is clear that $\delta_{jj,\text{RMA}}$ is also a function of $P$, but independent of the composition. This is only true in the isometric RMA limit. In the incompressible RMA limit, $\bar{v}_m = v_0$ and the solubility parameter is a constant.

However, as soon as mixing becomes non-isometric, the three average monomer volumes become different. In addition, the last two terms in Eqs. (40), and (35), respectively, do not vanish. This will usually give $\chi_{\text{eff,RMA}}$, and $c_{jj,\text{RMA}}$ and $\delta_{jj,\text{RMA}}$ some *complex* composition- and pressure-dependence. The situation becomes more complicated when we go beyond the RMA limit. In this case, we expect to find some unusual and complex behavior, as our numerical results will show below. The complex behavior of $\chi_{\text{eff}}$ for compressible and incompressible systems has been convincingly demonstrated earlier by Sariban and Binder[54] in their Monte Carlo simulation work, and by us in earlier publications.[12,14,47] The complex nature of $\chi_{\text{eff}}$ also implies a very complex nature of the cohesive energy density and solubility parameters. It is safe to conclude that these quantities are not *constants* as assumed in the regular solution theory.

**(iv) Scatchard-Hildebrand Conjecture.**

The Eq. (48) is related to the celebrated Scatchard-Hildebrand equation for the energy of mixing, obtained by multiplying $\chi_{\text{eff,RMA}}$ by $\phi_{m1}\phi_{m2}$; see Eq. (38). It is clear from Eq. (48) that this energy cannot be negative when all three bare chi's are positive. As we will see below, this will not be true in our theory. Indeed, our derivation of Eq. (48) shows that it works only in the isometric RMA limit. Thus, the Scatchard-Hildebrand equation cannot be justified in all cases. It should also be stated that one can define an effective chi $\hat{\chi}_{eff}$ by the use of volume ratios $v'$ and $v''$, instead of $\phi_{m1}$ and $\phi_{m2}$ in Eq. (38). It is easy to show that $\hat{\chi}_{eff} = \chi_{eff} \phi'_{m1} \phi''_{m2}$ and has no additional composition-dependence.

## IX. Gibbs Free Energy of Mixing

The adimensional Gibbs free energy of mixing is obtained by using Eqs. (19), and (21b). We have

$$\Delta_M \beta G = \sum_{j \geq 1} N_j \ln(V^{(j)}/V) + \sum_{j \geq 1} \left[ \left( N_{ju} \ln(\phi_{jj}/\phi_{jj}^{(j)}) - N_{ju}^{(j)} \ln(\phi_{jj}^{(j)0}/\phi_{jj}^0) \right) \right] \\
- \sum_{j \geq 1} (q/2) \left[ N \ln(\phi_{00}/\phi_{00}^0) - N^{(j)} \ln(\phi_{00}^{(j)}/\phi_{00}^{(j)0}) \right] - \sum_{j \geq 1} N_{mj} \ln(\phi_0/\phi_0^{(j)}) \quad . \quad (49)$$

The Gibbs free energy per monomer (particle) is

$$\Delta_M \beta g_m = \sum_{j \geq 1} \frac{y_j}{M_j} \ln(V^{(j)}/V) - \sum_{j \geq 1} y_j \ln(V_0 V^{(j)}/V V_0^{(j)}) \\
+ \sum_{j \geq 1} \left[ (\phi_{ju}/\phi_m) \ln(\phi_{jj}/\phi_{jj}^{(j)}) - (\phi_{ju}^{(j)}/\phi_m^{(j)}) \ln(\phi_{jj}^{(j)0}/\phi_{jj}^0) \right] \quad . \quad (50) \\
- \sum_{j \geq 1} (q/2\phi_m) \left[ \ln(\phi_{00}/\phi_{00}^0) - v^{(j)} \ln(\phi_{00}^{(j)}/\phi_{00}^{(j)0}) \right]$$

Even though the above equations do not show explicit dependence on $\chi_{ij}$, the Gibbs free energies are obviously determined by them. Let us consider the Gibbs free energy per monomer, and its relationship with the energy per monomer:

$$\beta \Delta_M g_m \equiv \beta \Delta_M e_m - \Delta_M s_m + \beta P \Delta_M v_m . \quad (51)$$

They differ in the last two quantities on the right hand side. Of the two, it is the entropy of mixing that is dominant and accounts for their major difference. Thus, if we want to use $\beta \Delta_M g_m$ as an estimator of $\chi_{12}$, we must correct for the entropy of mixing contribution in some way. Indeed, the second derivative of $\beta \Delta_M g_m$ with respect to $y$ has been used to estimate $\chi_{12}$ in scattering experiments,[55,14] since the derivative

$$(\partial^2 g_m / \partial y^2)_{T,P} \equiv (\partial^2 \Delta_M g_m / \partial y^2)_{T,P} \equiv (\partial \Delta \mu / \partial y)_{T,P} \quad (52)$$

is related to monomer number fluctuations.[55,56] Here,

$$\Delta \mu \equiv \mu_{m2} - \mu_{m1} . \quad (53)$$

More recently, we have carefully investigated this problem,[14,56] where we have focused on monomer number fluctuations in an ensemble, called the A-ensemble, in which $T$, $N$ or $V$, and $N_0$ or $V_0$ are kept fixed. This ensemble is a trivial generalization of the incompressible ensemble ($V_0 = 0$) to a fixed but non-zero free volume ensemble. It was shown that the following quantity

$$\Gamma_A(\chi_{12}, \chi_{01}, \chi_{02}) \equiv (\beta/2)(\partial \Delta \mu / \partial y)_{T,\phi_0} /(1-\phi_0)$$

can be used as an estimator of $\chi_{NR}$ in many cases. The origin of $(1-\phi_0)$ here is the same as in Eq. (42). The estimator is defined by the difference

$$\chi_{scatt}^{(A)} \equiv \Gamma_A(0,0,0) - \Gamma_A(\chi_{12}, \chi_{01}, \chi_{02}).$$

The difference in the two $\Gamma_A$'s primarily takes care of the entropy of mixing contribution, whose second derivative near $y = 0,1$ diverges; see Eq. (27). Since there is no density fluctuation in the A-ensemble, the number fluctuations are due to concentration fluctuations. In the grand canonical ensemble, to be called the C-ensemble in which $P$ replaces the free volume density $\phi_0$, we have both the density and the composition fluctuations that are always coupled,[56] despite a contrary claim in Ref. 55. We define a new effective chi in this ensemble via

$$\chi_{scatt}^{(C)} \equiv \Gamma_C(0,0,0) - \Gamma_C(\chi_{12}, \chi_{01}, \chi_{02}),$$

where

$$\Gamma_C(\chi_{12}, \chi_{01}, \chi_{02}) \equiv (\beta/2)(\partial \Delta\mu / \partial y)_{T,P} /(1-\phi_0).$$

The definitions are to maintain a parallel with the A-ensemble. However, we have shown elsewhere[51] that the following serves as a much better estimator of $\chi_{NR}$:

$$\overline{\chi}_{scatt}^{(A)} \equiv \Gamma_A(0, \chi_{01}, \chi_{02}) - \Gamma_A(\chi_{12}, \chi_{01}, \chi_{02}).$$

One can also introduce a similar quantity

$$\overline{\chi}_{scatt}^{(C)} \equiv \Gamma_C(0, \chi_{01}, \chi_{02}) - \Gamma_C(\chi_{12}, \chi_{01}, \chi_{02})$$

in the C-ensemble. However, it still contains contributions from density fluctuations, and does not reliably estimate $\chi_{NR}$ in all cases.

It is easy to show[51] that

$$\Gamma_A = \Gamma_C + (\overline{v}_1 - \overline{v}_2)^2 / 2TK_T,$$

where $\overline{v}_j$ are the partial monomer volumes, and $K_T$ is the isothermal compressibility. Thus, we can extract $\chi_{scatt}^{(A)}$, and $\overline{\chi}_{scatt}^{(A)}$ from $\chi_{scatt}^{(C)}$, and $\overline{\chi}_{scatt}^{(C)}$, respectively.

## X. Numerical Results

The results presented here are based on a recent work,[57] which contains additional results not presented here. We restrict most of our analysis to $w_{12} = 0.9975$. This corresponds to a very small $\chi_{12} \cong 0.02$, and $\chi_{NR} \cong 0.015$ for $q = 8$. As discussed in Ref. 14, $v_0 = 20 \pm 2$ Å$^3$ and the range (0.6-0.8) for $w_{01}$ and $w_{02}$ seem realistic. Here, we will take $w_{01}$ and $w_{02}$ to be 0.75 and 0.76. We also consider $w_0 \equiv w_{01} = w_{02} = 1.0$ and 0.95 for comparison with earlier results,[12] where $w_0$ denotes the common values of $w_{01}$, and $w_{02}$ when they are equal. At room temperatures and using the above value of $v_0$, we find that the adimensional pressure $z_0 \equiv \beta P v_0 = 0.2$ represents the atmospheric pressure. Thus, we mostly consider $z_0 = 0.20$ and 1.0. In most cases, we consider two different blends. In the one blend, we take $M = M_1 = M_2$ equal to 100 and 1,000. (We will always use $M$ to denote the DP, whenever, we have the two DP's equal: $M_1 = M_2$.) In the other, we take $M_1 = 1{,}000$, and $M_2 = 100$. However, a few other cases are also investigated. We will use the aspect ratio $a \equiv M_1/M_2$ as a measure of the asymmetry or the size-disparity in DP's.

**(i)** **Effect of $q$**

As soon as we leave the RMA limit such as when $q$ is finite, the entropy of mixing *cannot* be put in a form suitable of continuum interpretation. We have already shown, see Fig. 3 in Ref. 46, that the contact densities are not simply proportional to $q$. Moreover, they are complicated functions of $\varepsilon_{ij}$. Hence, $\Delta_M s_{m,unb}$ (●,▼,■) and $\Delta_M s_{m,cont}$ (○, ▽, □) will also change with $q$, as we show in Fig. 4(a). The results in Fig. 4 are for $z_0 = \beta P v_0 = 0.2$, $M = 1000$, $w_{12} = 0.985$, $w_{01} = 0.75$, and $w_{02} = 0.76$. We have taken three different values 8(●,○), 16(▼,▽) and 24(■,□) for $q$. We see that $\Delta_M s_{m,unb}$ and $\Delta_M s_{m,cont}$ are *negative* as expected and are strongly influenced by $q$, but the effect is mostly *opposite*. While $\Delta_M s_{m,unb}$ decreases in magnitude as $q$ increases and almost vanishes for $q = 24$, $\Delta_M s_{m,cont}$ continues to increase in magnitude in the midrange. However, near $y = 0$ or 1, it decreases in magnitude. In Fig. 4(b), we show the effect of $q$ on $\Delta_M s_m$; it decreases rapidly as $q$ increases. From the behavior of $\Delta_M s_{m,cont}$ in Fig. 4(a), we conclude that $\Delta_M s_{m,cont}$ mostly controls the behavior of $\Delta_M s_m$ for large $q$. The effect of $q$ on $\beta \Delta_M e_m$ is shown in Fig. 4(c) for $q = 8$ (●), $q = 16$ (○) and $q = 24$ (▼). As expected, $\beta \Delta_M e_m$ increases with $q$ for fixed $\beta$ due to the increase in the contact densities. Since the coordination number has no continuum analog, the above effect is purely due to a lattice structure. However, decreasing $\beta$ as $q$ increases such that $\beta q$ remains fixed and finite gives us the RMA limit in Eq. (37).

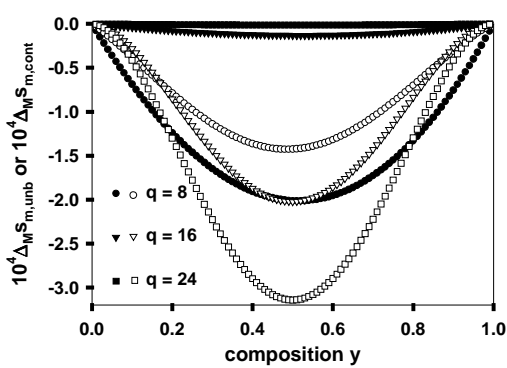

(a)

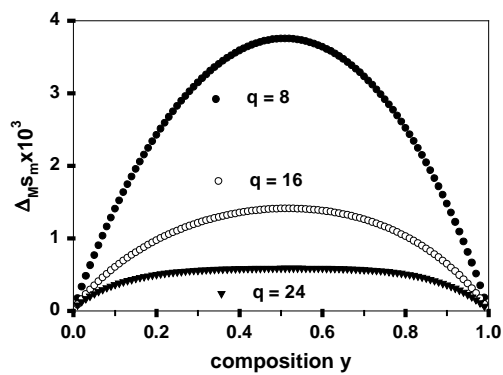

(b)

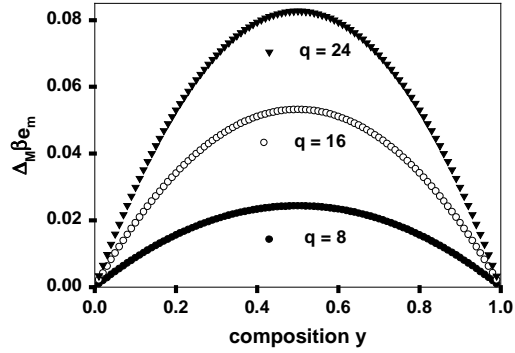

(c)

Fig. 4. The effect of $q$ on (a) $\Delta_M s_{m,unb}$ (●,▼,■) and $\Delta_M s_{m,cont}$ (○,▽,□); (b) $\Delta_M s_m$, and (c) $\Delta_M \beta e_m$. We take a blend with $M = 1000$, $w_{12} = 0.985$, $w_{01} = 0.75$, $w_{02} = 0.76$ and $z_0 = 0.2$.

### (ii) Effect of $w_{01}$ and $w_{02}$

We now fix $q = 8$ for the remaining results. We consider the effect of the pure component interactions described by $w_{01}$, and $w_{012}$ on $\Delta_M v_m$, $\Delta_M s_m$, $\beta\Delta_M e_m$, and $\chi_{eff}$ in Figs. 5(a-d), respectively. We fix the composition at $y = 1/2$, and take $z_0 = 0.2$, $w_{02} = cw_{01}$, where $c = 1.012$, and 1, and $M = 100$. We have found[57] that the free volume density $\phi_0$ (result not shown here) increases monotonically, while the above quantities show a *maximum* and/or a *minimum* as a function of $w_{01}$, depending on the value of $w_{12}$.

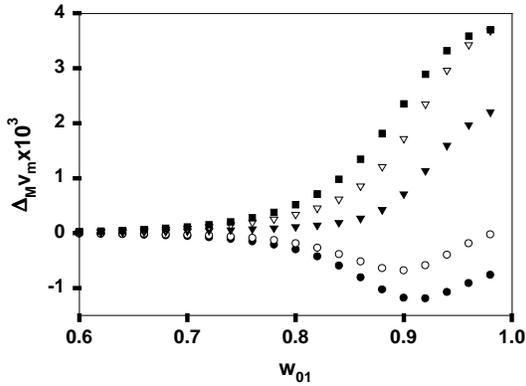

(a)

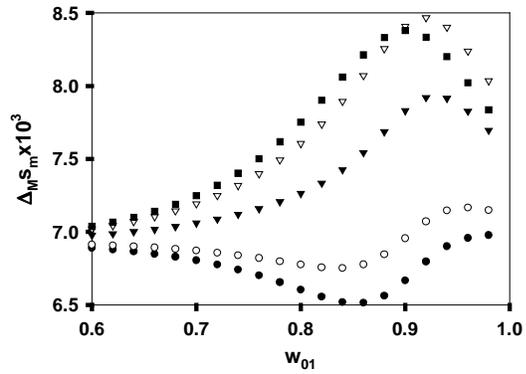

(b)

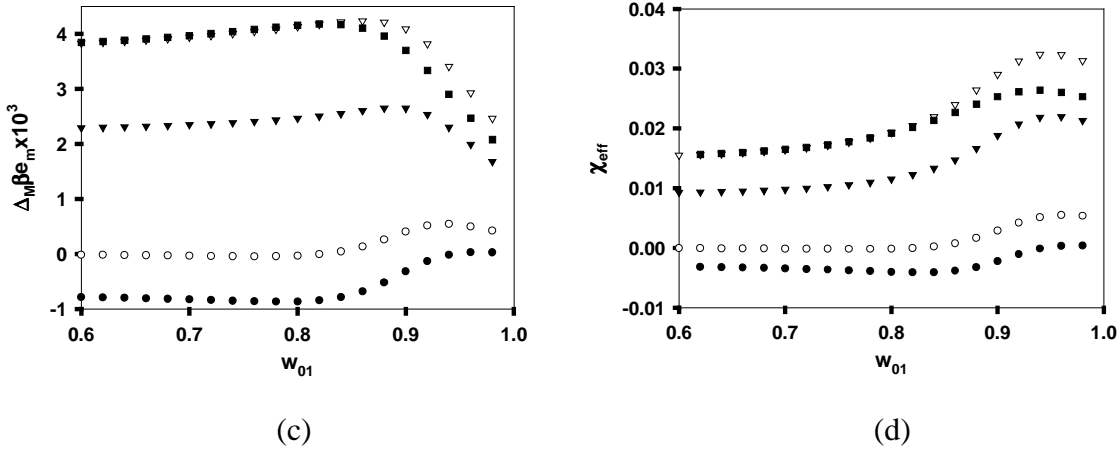

(c)                                                        (d)

Fig. 5. Effect of $w_{01}$ and $w_{02} = cw_{01}$ on (a) $\phi_0$, (b) $\Delta_M v_m$, (c) $\Delta_M s_m$, (d) $\Delta_M \beta e_m$ and (e) $\chi_{eff}$. We have a blend with $M = 100$, $q = 8$, $y = 0.5$ and $z_0 = 0.2$. $c = 1.012$ for $w_{12} = 1.005$ (●), 1.00 (○), 0.9985 (▼), 0.9975 (▽). $c = 1.0$ for $w_{12} = 0.9975$ (■).

For the equal-DP case considered here, a negative $\Delta_M v_m$, Fig. 5 (a), signals that there is effectively an *attractive interaction* between polymers when they are mixed, even though the values of $w_{12}$ imply a repulsive interaction. The effective attraction comes about due to an interaction-asymmetry between $w_{01}$ and $w_{02}$. (For a symmetric blend, $\Delta_M v_m$ *cannot* be negative for repulsive $w_{12}$, as said in Sect. VI.) Thus, we may also obtain a *negative $\beta\Delta_M e_m$* in these cases. This is seen clearly in Fig. 5(c). The energy of mixing is negative for $w_{12} = 1.0$ (○) for $w_{01} \lesssim 0.8$, though it is not so evident in the figure; see Fig. 6 below. The entropy of mixing $\Delta_M s_m$ also exhibits (●, ○) a minimum in these cases, as shown in Fig. 5(b). We show $\chi_{eff}$ in Fig. 5(d), where we see both positive and negative values. Positive values of $\chi_{eff}$ vary by a factor of two (▼, ▽, ■). In all cases, the values of $\chi_{eff}$ remain close to their respective $\chi_{NR}$ values of about 0.015 (▽, ■), 0.009 (▼), 0.0 (○), and –0.003 (●) for $w_{01} \lesssim 0.8$. For larger values of $w_{01}$, we begin to see strong deviations from their approximate equality. The peaks or valleys in Fig. 5 are close to $w_{01} = 1$ and occur at different values of $w_{01}$ for different quantities. The non-monotonic behavior in $w_{01}$ will be seen in many results presented below. The void density is monotonic increasing as a function of $w_{01}$ in all cases we have considered.

It is easy to understand some of the features seen in Fig. 5. At $w_{01} = 1$, the behavior of various quantities can be understood from our earlier investigation.[12] As $w_{01}$, and $w_{02}$ vanish, we obtain an incompressible mixture and pure components and $\Delta_M v_m = 0$. (There is coexistence in this limit between the incompressible blend and a pure vacuum; however, we are not interested in investigating coexistence here, which has been studied in Ref. 13.) Hence, the volume of mixing must be zero. If $\Delta_M v_m$ becomes negative at some intermediate $w_{01}$, then there must exist a minimum in $\Delta_M v_m$ as a function of $w_{01}$ since it vanishes at $w_{01} = 0$. Thus, the minimum in $\Delta_M v_m$ is a *necessity*. The relative size of the two terms in Eq. (36) determines whether $\beta\Delta_M e_m$ has a peak or becomes negative. The origin of a negative $\beta\Delta_M e_m$ is in the first sum. The other sum is positive for repulsive interactions. It is clear that for the case $w_{02} = 1.012\, w_{01}$ (○), it will take a value of $w_{12}$ strictly less than 1 to make $\beta\Delta_M e_m$ positive. We find a negative $\chi_{eff}$ (○) for $w_{12} = 1.00$.

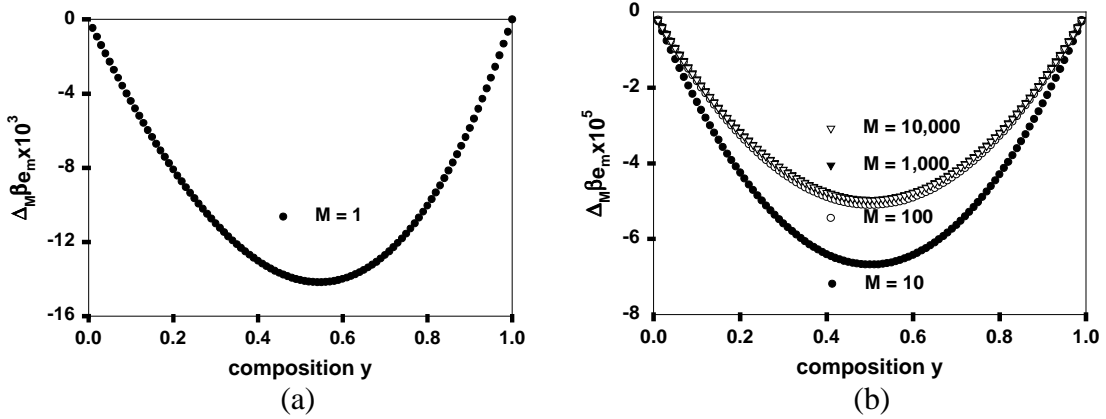

Fig. 6. Violation of Scatchard-Hildebrand conjecture and negative energy of mixing for repulsive interactions for (a) $M = 1$ (●), (b) $M = 10$ (●), $10^2$ (○), $10^3$ (▼), $10^4$ (▽). We take $q = 8$, $w_{12} = 1.0$, $w_{01} = 0.75$, $w_{02} = 0.76$ and $z_0 = 0.2$.

**(ii)    Scatchard-Hildebrand Conjecture**

An *important* consequence of a negative $\beta\Delta_M e_m$ when all the interactions are *repulsive* is that the conventional Scatchard-Hildebrand equation[2] [see Eqs. (38, 48), and the discussion thereafter], according to which the energy of mixing for non-polar mixtures *cannot* be negative, is generally *not* valid. A similar conclusion was drawn recently when we considered the effective chi measured in small angle neutron scattering.[14] For $w_{12} = 1$, $\chi_{12} = 0$ and the contribution to $\beta\Delta_M e_m$ comes from the first term in Eq. (36). We show the results for $\beta\Delta_M e_m$ for different values of $M$ from 1 to $10^4$ in Figs. 6(a, b). It is negative and decreases in magnitude as $M$ increases and almost reaches its asymptotic value for $M \geq 10^4$. For asymmetric blends, the situation is more complicated.[57] In this case, the values of $\beta\Delta_M e_m$ depend on the aspect ratio $a$. We do not show the results here, but summarize them below. For a given $a$, $\beta\Delta_M e_m$ decreases in magnitude as the DP $M_1$ increases. For a given $M_2$, it increases in magnitude as $a$ increases. The violation is stronger for small DP's, the case relevant for simple fluids. The negative contribution is due to the free volume contribution in our theory and is *not* accounted for in the classical Scatchard-Hildebrand theory.

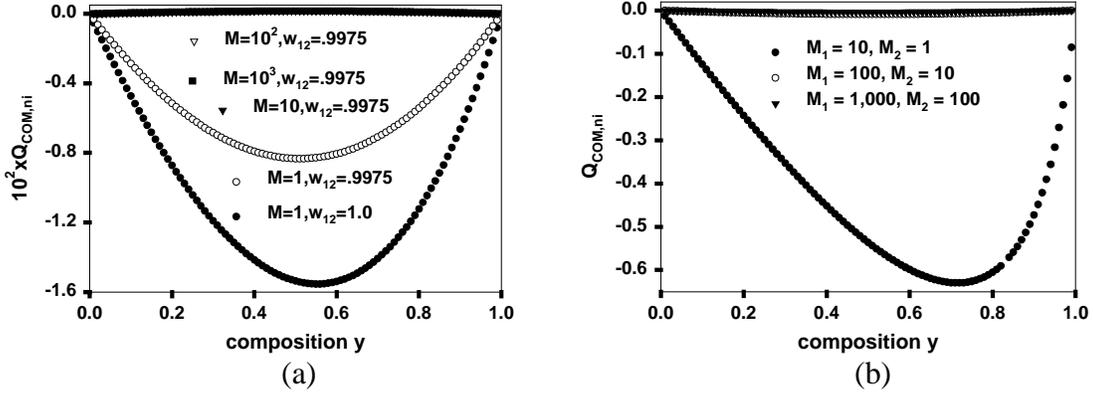

Fig. 7. Non-ideal contribution to the center-of-mass entropy of mixing in asymmetric blend with $q = 8$, $w_{01} = 0.75$, $w_{02} = 0.76$ and $z_0 = 0.2$. (a) For $w_{12} = 1$, $M = 1$ (●); for $w_{12} = 0.9975$, $M = 1$ (○), 10 (▼), $10^2$ (▽), $10^3$ (■). (b) For $w_{12} = 0.9975$, $M_1, M_2 = 10, 1$ (●), $10^2, 10$ (○), $10^3, 10^2$ (▼).

### (iv) Non-Ideal COM Contributions

The contribution $\Delta_M s_{m,\text{COM}}$ is a function of volume ratios $v'$ and $v''$ and not $y_1$ and $y_2$, when the mixing is not isometric. The "ideal" entropy of mixing is $-(y_1/M_1)\ln y_1 - (y_2/M_2)\ln y_2$. The remainder of the COM part is the non-ideal contribution, and is given by

$$\Delta_M s_{m,\text{COM,ni}} = -(y_1/M_1)\ln(v'_m/v_m) - (y_2/M_2)\ln(v''_m/v_m).$$

In Fig. 7, we plot $Q_{\text{COM,ni}} \equiv \sqrt{M_1 M_2}\,\Delta_M s_{m,\text{COM,ni}}$ for different values of $M_1$ and $M_2$, but fixed $a$. We fix $z_0 = 0.2$. We note that $Q_{\text{COM,ni}}$ is negative and large in magnitude for the smallest DP, but very small and positive for larger DP's. It reaches an asymptotic form for larger DP's rapidly. The disparity between $w_{01}$ and $w_{02}$ for the smallest DP $M = 1$ (●, ○) is strong enough to make $Q_{\text{COM,ni}}$ negative, but becomes weak enough for larger DP's and makes $Q_{\text{COM,ni}}$ small and positive. Thus, a strong asymmetry produces a negative $Q_{\text{COM,ni}}$; see Fig. 7(b). The contribution remains negative but is relatively large compared to its symmetric counterpart in Fig. 7(a). Nevertheless, it contribution is negligibly *small* for large DP's.

### (v) Non-Ideal Contributions in $\Delta_M s_m$

For $M = M_1 = M_2$, consider $Q_{\text{COM}} \equiv M\Delta_M s_{m,\text{COM}}$ and $Q_S \equiv M\Delta_M s_m$, which are equal and independent of $M$ in a regular or ideal solution theory. This is not true in our theory. While $Q_{\text{COM}}$ has *no* noticeable dependence on $M$ (results not shown), $Q_S$ exhibits a very *strong* dependence on $M$ as shown in Figs. 8(a, b). The rate of increase becomes more pronounced with $M$. The strong dependence in $Q_S$ but almost unnoticeable dependence in $Q_{\text{COM}}$ comes from the remaining three *non-ideal* or *non-random* contributions to $\Delta_M s_m$.

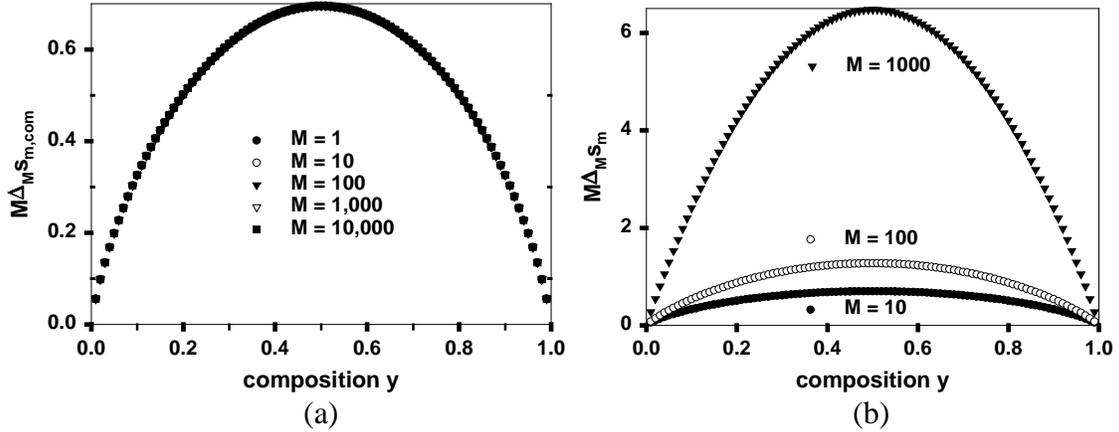

Fig. 8. Effect of $M$ on the entropy of mixing of a symmetric blend with $q = 8$, $w_{12} = 0.9975$, $w_0 = 1.0$, and $z_0 = 0.1$.

### (vi) Effect of $w_0$ and $z_0$

Below we summarize some numerical findings without reproducing the results, which are given in Ref. 57. We consider two blends with DP's 100/100 and 1000/100 with $w_{12} = 0.9975$. We consider various values of $w_0 \equiv w_{01} = w_{02}$, and two values of $z_0 = 0.2$ and 1.0. We find that $\Delta_M v_m > 0$, and behaves the same way for both blends. As $w_0$ decreases, the volume change decreases. This is consistent with Fig. 5(b) where $\Delta_M v_m$ (■) is monotonic increasing function of $w_0$ for $w_{12} = 0.9975$. The decrease in $\Delta_M v_m$ as $w_0$ decreasing is related to the fact that the free volume $\phi_0$ decreases rapidly. The amount of free volume for $w_0 \cong 0.75$ is small (~2-5%), and this makes the blend a *dense* liquid. Near $w_0 \cong 1.0$, studied in Ref. 12, the blend is in a gaseous state since the free volume is very large (~50%). The effective chi depends on the DP's. Apart from this difference, the two blends have similar features for $\chi_{eff}$. In both blends, the magnitude of $\chi_{eff}$ increases as the pressure decreases. The same feature was observed earlier[12] for $w_0 = 1$. However, the dependence of $\chi_{eff}$ on $w_0$ is non-monotonic due to non-monotonic nature of the energy of mixing, as seen in Fig. 5(c). The value of $\chi_{eff}$ first increases and then decreases so that its value for $w_0 = 1$ lies between its value for $w_0 = 0.95$ which is higher and for $w_0 = 0.75$ which is lower. We also find that the symmetric blend has a higher $\chi_{eff}$ than the asymmetric blend. We have also seen a *negative* curvature in $\chi_{eff}$.[57] The complex behavior of $\chi_{eff}$ implies complex dependence in the cohesive energy density and the solubility parameter.

## XI. Conclusion and Summary

The present review deals with the statistical mechanics of a lattice model of a multi-component system, since the lattice model is devoid of certain inherent limitation of continuum models as discussed here. On a lattice, the free volume is modeled by voids.[15,16] Voids and monomers have the same volume in this basic lattice formulation.

As we discuss below, this condition can be relaxed by making the model little more complex, but without altering the general approach given here. We have argued that, since voids determine the pressure of the system, they play a very different role than material species. In particular, voids do not form a pure component by themselves. This has not always been the case.[39,40] The partition function of the model is defined in terms of microscopic or bare parameters, their values being of central importance in the thermodynamics of the system. One of these parameters is the bare chi parameter whose measurement has been of vital interest. It has been proved that the bare parameters appearing in the partition function must be independent of the thermodynamic state of the system, and of the properties of the lattice. Otherwise, there would result thermodynamic violations. This aspect of the bare parameters has been misunderstood in the literature. We show that the voids that are used to introduce free volume in the model must be carefully treated, and the mixing functions must be carefully defined. Again, this aspect has not been correctly appreciated in the literature.

An exact theory of the model can allow for the extraction of the exact values of the bare parameters from accurately measured thermodynamic properties. Unfortunately, it is a fact that we have no exact theory at present. Hence, what one extracts from experiment is not the bare quantities, but some effective or dressed quantities that must depend on the thermodynamic state of the system.

We have argued that simple-minded continuum theories treat particles as point-like. The *ideal entropy of mixing* considered in ideal and regular solution theories can be justified only for *point-like* particles, for which mixing is isometric. It is in this case that the lattice and continuum theories appear to have the same form of the entropy of mixing. When particles have a size, the situation becomes complicated. The simple formulation due to van der Waals and Hildebrand[6,25] to incorporate finite size particles also *cannot* be called successful because they do not give rise to a volume change on athermal mixing for particles of *different* sizes. We trace the problem to the fact that both theories contain only free volumes but not total volumes. The pressure in lattice theories diverges logarithmically, but as a power law in continuum theories. One of our interests in the review has been to investigate whether thermodynamic functions in the lattice theory can be cast in a form that remains meaningful in continuum. We are not interested in matching the behavior of the pressure. For the latter, we refer to Ref. 58. As we conclude in the review, our final conclusion is negative as soon as we go beyond the random-mixing approximation. This should not come as a surprise since the lattice coordination number $q$ has no continuum analog. It should be noted that the lattice and continuum equations of state for non-zero size particles are quite different, as already noted.[12]

After this general discussion, which should be valid in any lattice theory, we focus our attention on our recent lattice theory of compressible polymer mixture to calculate various mixing functions at constant temperature and pressure. Some of these functions have not been properly identified in the literature. For the sake of clarity, we restrict ourselves to linear chains but the theory is applicable to polymers of *any* fixed architectures. Our theory, which supports a non-zero volume of mixing for asymmetric blends, contains both kinds of the volumes explicitly.

For finite $q$, our lattice theory contains contributions that have no continuum analog. In particular, the contribution from chemically unbonded lattice bonds (due to polymer connectivity) and contact bonds *cannot* be put in a form suitable for continuum description. We calculate various mixing quantities for a multi-component mixture. Since our theory is based on non-random mixing, it contains corrections to the ideal and regular solution theories, which are important in many situations, as we have demonstrated here. We show that the two currently used extensions of the FH-theory for a compressible binary mixture are not appropriate. We remedy this shortcoming. The *non-ideal* contributions due to free volume and due to chemically unbonded bonds in Eqs. (28), and (31), respectively, are non-zero whenever there is volume of mixing. We show that the ideal $\Delta_M S$ is valid for only athermal mixtures of particles of identical size because of isometric mixing. For particles of different sizes, there is in general a non-zero volume of mixing.

The COM contribution to the entropy in our theory is exactly the ideal entropy of mixing for isometric mixing. A correction is needed for non-isometric mixing, which is minor for large DP's. For small DP's that are suitable for simple fluids, the correction is large. The most important non-ideal correction is from the free volume contribution in Eq. (28), which is mostly insensitive to the polymer DP's and eventually controls the entropy of mixing. The entropy of mixing has a DP-dependence very *different* from that seen in the COM entropy of mixing.

In general, volume of mixing has many important consequences. In some cases, it is responsible for the violation of Scatchard-Hildebrand theory of energy of mixing, since non-isometry can give rise to a negative energy of mixing even though all interactions are repulsive. The violation is due to the contribution from the first term in Eq. (36). It also causes the cohesive energy density and the solubility parameter to have a composition- and pressure-dependence, which is missing in a regular solution theory. These quantities are constant only in an isometric RMA. Non-isometry brings about variations from the regular solution theory. We also provide the RMA limit of our theory for a compressible mixture, which should be used in place of the original F-H theory. The resulting equation of state in Eq. (39) is different from the Sanchez-Lacombe equation.[30,35,36] The calculation of mixing functions in the latter theory is shown to have a few internal inconsistencies. The study provides us with new insight and corrects some misconceptions as discussed above in this section and elaborated in previous sections.

Finally, we wish to add some remarks about the possible extension of our lattice model to incorporate cases in which monomers of various species occupy different volumes.[59] We look for the largest cell volume $v_0$, so that the volumes occupied by monomers of each species are integer (or close to integer) multiples of $v_0$. This will also make the void volume equal to $v_0$, which is desirable, as the voids are not supposed to have an intrinsic size by themselves (unless we are in a regime where quantum field effects become important, which we are going to neglect here). Moreover, it is also undesirable to have a void occupy the same amount of volume as a bulky monomer. If anything, it must occupy very small volume. Our strategy ensures that voids have the smallest volume of all types of monomers on the lattice. Indeed, if the need be, one can choose even a smaller volume obtained by dividing $v_0$ by an integer of our choice. Once we make the choice, we call that volume our elemental cell volume $v_0$. Each *j*-th species monomer occupies an integer number $m_j$ of lattice sites, so that the monomer volume $v_j \cong m_j v_0$. In order to ensure that the $m_j$ sites represent somewhat of a compact monomer, we take the $m_j$ sites to form a compact "dendrimer"[49] on the Bethe lattice on which our lattice theory is obtained. It should be recalled that our lattice theory is an exact solution of our model on a Bethe lattice.[49]

In summary, we have shown that a finite $q$ lattice theory cannot be put in a form valid for a continuum version. We have also shown that the van der Waals and Hildebrand equations are not appropriate for mixtures of particles of *different* sizes. We have calculated various mixing functions at constant temperature and pressure within the framework of our recently developed theory of compressible polymer mixture. We show that the entropy of mixing is different from its center-of-mass part, the latter being very similar but not identical to the ideal entropy of mixing. We show that the volume of mixing, the interactions and the finite coordination number of the lattice are the cause of most of the non-ideal contributions for a compressible mixture. We give some numerical results using our recent lattice theory of multi-component polymer mixtures.

Finally, I am thankful to Andrea Corsi for help with some figures, and with reading.

# Appendix I

Let us, for simplicity, consider the canonical ensemble for an incompressible blend. Let $N_c$ denote the number of nearest-neighbor contacts between the two species, and let $w = \exp(-\beta\varepsilon)$. The corresponding partition function is given by

$$Z_N \equiv \sum_{\Gamma:N_c} w^{N_c} ; \qquad (A.1.1)$$

the sum is over all possible values of $N_c$, consistent with a lattice of $N$ sites. The Helmholtz free energy per site is given by $f_N = -(T/N)\ln Z_N$ for finite $N$. In the thermodynamic limit $N \to \infty$, we expect the sequence $\{f_N\}$ to converge to the thermodynamic limit $f$, which must satisfy the thermodynamic relation

$$(\partial f / \partial T) = -s . \qquad (A.1.2)$$

Let us calculate this derivative for fixed $N$, and then take the limit. We find that

$$(\partial f_N / \partial T) = -(1/N)\ln Z_N - (T/NZ_N)\sum w^{N_c}[\varepsilon N_c / T^2 - \beta\varepsilon' N_c],$$

where $\varepsilon' = (\partial\varepsilon/\partial T)$. Noting that $e_N = (\varepsilon/NZ_N)\sum N_c w^{N_c}$, we find that

$$(\partial f_N / \partial T) = f_N/T - e_N(1/T + \varepsilon'/\varepsilon) = -s_N + e_N \varepsilon'/\varepsilon .$$

In the thermodynamic limit, we get the limiting equation

$$(\partial f / \partial T) = -s + e\varepsilon'/\varepsilon , \qquad (A.1.3)$$

which differs from Eq. (A.1.2) by a term controlled by the derivative $\varepsilon'$. Thus, this derivative must vanish identically, i.e. $\varepsilon$ must be independent of the temperature if we want to maintain the thermodynamic relation (A.1.2). The preceding argument can be extended to show that there should not be any *P*-dependence also.

## Appendix II

Consider, for example, the thermodynamic relation

$$\beta\mu_{mj} \equiv -(\partial S/\partial N_{mj})_{V,\bar{N}_j,\{N_{ij}\}}, \qquad (A.2.1)$$

where $\bar{N}_j$ denotes the set of all other particle numbers besides $N_j$, and $\{N_{ij}\}$ the set of all contact densities. Of course, we keep the set $\{M_j\}$ also fixed in the derivative. We now show that this relation is violated for composition-dependent $v_0$. Suppressing the sets $\bar{N}_j$, and $\{N_{ij}\}$ in the derivative, we find that

$$(\partial S/\partial N_{mj})_V = (\partial S/\partial N)_{N_{mj}}(\partial N/\partial N_{mj})_V + (\partial S/\partial N_{mj})_N \qquad (A.2.2)$$

by treating $V$ as a function of $N$, and the composition. In a lattice theory, $S$ is a function of $N$, and not $V$, as discussed in Sect. IV. According to Eq. (11b), the last term in Eq. (A.2.2) represents the adimensional chemical potential $(-\beta\mu_{mj})$. Thus, the first term above describes the violation of the above thermodynamic relation (A.2.1). The discrepancy is determined by the derivative $(\partial v_0/\partial N_{mj})_V$ and vanishes for composition-independent cell volume.

## Appendix III

Ryu[50] derived Eq. (24a) for a binary mixture. The relation and its proof are easily extended to a multi-component system, as we show below.

From the definition, see Sect. IV (b), and Eq. (2.4) in Ref. 41, we note that

$$\phi_{ju} = \phi_{jj} + \sum_{k>j}\phi_{jk} + \sum_{k<j}\phi_{kj}, \qquad j \geq 0. \qquad (A.3.1)$$

Using Eqs. (18) we can rewrite (A.3.1) as follows:

$$\phi_{ju} + (1/2)\left(\sum_{k>j}\bar{w}_{jk}\phi_{jk} + \sum_{k<j}\bar{w}_{kj}\phi_{kj}\right) = \sqrt{\phi_{jj}}\sigma, \qquad j \geq 0, \qquad (A.3.2)$$

where we have introduced the combination

$$\sigma \equiv \sum_{j\geq 0}\sqrt{\phi_{jj}},$$

and $\bar{w}_{ij} = 1/w_{ij} - 1$. Summing all the equations in (A.3.2), and noting that $\phi_u = \sum_{j\geq 0}\phi_{ju}$, we find

$$\sigma^2 = D \equiv \phi_u + \sum_{0\leq i<j}\bar{w}_{ij}\phi_{ij}. \qquad (A.3.3)$$

Using $\bar{D}_j$ to denote the left-hand side of (A.3.2), we finally get the solution for $\phi_{jj}$, by rewriting it as follows:

$$\phi_{jj} = \bar{D}_j^2/D. \qquad (A.3.4)$$

This is the generalization of Ryu's set of equations to a multi-component mixture.